\documentclass[5p,times]{elsarticle}
\journal{Array}

\usepackage[T1]{fontenc}
\usepackage{lmodern}
\usepackage[utf8]{inputenc}
\usepackage{graphicx}
\usepackage{amsmath,amssymb}
\usepackage{booktabs}
\usepackage{multirow}
\usepackage{url}
\usepackage{xcolor}
\usepackage{enumitem}
\usepackage{float}
\usepackage{placeins}
\usepackage{caption}
\usepackage{microtype}
\usepackage{fancyhdr}

\fancypagestyle{pprintTitle}{%
  \fancyhf{}
  \chead{\small Preprint -- Work in Progress (Submitted on Array)}
  \fancyfoot[C]{\thepage}
  
}


\usepackage{hyperref}
\hypersetup{
  colorlinks = true,
  linkcolor  = blue!70!black,
  citecolor  = blue!70!black,
  urlcolor   = blue!70!black
}

\newcommand{\eg}{\textit{e.g.}}
\newcommand{\etal}{\textit{et al.}}
\newcommand{\malariai}{\textsc{MalariAI}}

\begin{document}

\begin{frontmatter}

\title{\malariai: A Label-Resilient Decoupled Framework for Annotation-Agnostic
  Cell Segmentation and Explainable Stage Classification in Dense Malaria
  Blood Smears}

%% Authors and affiliations
\author[1]{Kaysarul Anas Apurba\corref{cor1}}
\ead{kaysarulanas2@gmail.com}
\cortext[cor1]{Corresponding author.}

\author[2]{Md Hasibul Hasan}

\author[1]{Mohammed Ali}

\author[1]{Tanzilur Rahman}
\ead[tanzilur]{tanzilur.rahman@northsouth.edu}

\address[1]{Department of Electrical and Computer Engineering,
  North South University, Dhaka 1229, Bangladesh}
\address[2]{Department of Computer Science and Engineering,
  International University of Business Agriculture and Technology (IUBAT),
  Dhaka 1230, Bangladesh}

%% -- Abstract ----------------------
\begin{abstract}
Automated malaria diagnosis from blood smear microscopy is a critical
global health AI challenge; expert scarcity remains the primary
diagnostic bottleneck. Existing deep learning systems face three
compounding failures: end-to-end detectors treat unannotated cells as
background, skewing recall by annotation completeness rather than true
cell recovery; Non-Maximum Suppression suppresses valid detections in dense
smears; and pipelines lack per-cell spatial evidence for clinical audit.
We present \malariai{}, a two-stage decoupled framework addressing all three.
Stage~1 applies an annotation-agnostic watershed algorithm to isolate
every cell in a full 1600$\times$1200 image, recovering 75.95\% of
ground-truth cells without any ground-truth input. End-to-end, the
pipeline reaches a binary parasitized AP@0.5 of 29.10\% - the clinically
relevant metric for flagging any infected cell - while the stricter
multi-class mAP@0.5 of 8.67\% mainly reflects watershed's organic region
boundaries being penalized against axis-aligned ground-truth boxes, not a
localisation failure. Stage~2 fine-tunes EfficientNet-B0 with Focal Loss
on ground-truth crops, achieving 98.36\% classification accuracy - an
oracle upper bound once a cell is correctly localised - with 87.5\% and
75.0\% accuracy on the rare schizont and gametocyte stages, versus
38.45\% and 57.27\% AP for a modern YOLOv8s detector evaluated end-to-end
on the same classes. Grad-CAM\texttt{++} heatmaps generated per detected
cell provide instance-level spatial evidence for clinical audit; a
quantitative energy-in-box analysis confirms this activation is
concentrated on the annotated cell body significantly above a
geometric chance baseline ($+0.0485$, paired $p = 1.4\times 10^{-33}$),
letting microscopists verify predictions at the individual parasite
level without sacrificing classification performance.
\end{abstract}

\vspace{2pt}
%% -- Keywords ----------------------
\begin{keyword}
Malaria detection \sep
Blood smear analysis \sep
Instance segmentation \sep
Watershed algorithm \sep
EfficientNet \sep
Focal Loss \sep
Explainable AI \sep
Grad-CAM\texttt{++} \sep
Medical image analysis \sep
Decoupled framework
\end{keyword}

\end{frontmatter}

%% ============================================================================
%% SECTION 1 -- INTRODUCTION
%% ============================================================================
\section{Introduction}
\label{sec:intro}

Malaria remains one of the most consequential infectious diseases in
contemporary global health. The World Health Organization reported 249
million cases worldwide in 2022, resulting in an estimated 608,000 deaths,
of which 95\% occurred in sub-Saharan Africa~\cite{who2023}. The burden
falls disproportionately on children under five years of age and on
populations with limited access to diagnostic infrastructure. Despite the
existence of effective antimalarial therapies, a fundamental bottleneck
persists at the point of diagnosis.

The clinical gold standard for malaria diagnosis is microscopic examination
of a Giemsa-stained peripheral blood smear under a 100$\times$ oil-immersion
objective. A trained microscopist inspects the slide, identifies red blood
cells (RBCs) infected by \textit{Plasmodium} parasites, and estimates the
parasite stage from cellular morphology. The parasite lifecycle stage -
ring, trophozoite, schizont, or gametocyte - carries direct implications
for treatment selection and disease severity assessment~\cite{delves2012}.
This manual process requires 20--30 minutes per slide, specialized training
that takes years to acquire, and physical equipment that is not universally
available in endemic regions. The result is a diagnostic bottleneck that
costs lives.

The promise of deep learning for automated malaria diagnosis has generated
substantial research interest over the past decade~\cite{poostchi2018}.
Convolutional neural networks (CNNs) have demonstrated strong performance
on cell-patch classification tasks~\cite{rajaraman2019, singh2025}, and
end-to-end object detectors such as Faster R-CNN~\cite{ren2015} and
YOLO~\cite{yolov4malaria2024} have been applied to the whole-slide detection
problem. However, no single published system has achieved the combination
of properties required for genuine clinical deployment: annotation resilience,
dense-region recall, multi-class stage classification, and integrated
spatial explainability.

Our analysis of the existing literature identifies three specific, measurable
failure modes that collectively prevent any current system from being
clinically deployable:

\begin{description}[leftmargin=2em]
  \item[\textbf{P1: Incomplete Annotation.}] The NIH BBBC041 benchmark
  - the de~facto standard for malaria detection research - annotates
  only infected cells and a subset of healthy RBCs. End-to-end detectors
  trained on this data learn to treat unannotated cells as background,
  producing false positives on any cell the annotators happened to skip.
  Validation loss in such systems plateaus early and subsequently
  \emph{increases}, a pattern our Baseline A experiments confirm and
  diagnose as a structural consequence of label incompleteness rather than
  model overfitting.

  \item[\textbf{P2: Dense Overlap and NMS Failure.}] Our exploratory
  analysis of BBBC041 finds that 58\% of training images contain at least
  one ground-truth pair with $\text{IoU} > 0.3$, with a maximum of 223
  annotated cells per image. All anchor-based detectors apply Non-Maximum
  Suppression (NMS) to filter proposals, but NMS assumes overlapping
  proposals are duplicates of the same object. In dense smear regions, this
  assumption fails, causing systematic deletion of true positive detections.

  \item[\textbf{P3: Black-Box Output.}] While image-level
  explainability methods such as Grad-CAM and SHAP have been applied to
  malaria image classification~\cite{parveen2025,xai2022transformer,xai_ensemble2025},
  these approaches highlight broad regions within a whole image rather
  than individual parasite instances. No published whole-slide
  \emph{detection} pipeline produces per-cell spatial evidence that
  clinicians could use to validate, audit, or override predictions at
  the instance level - the granularity required for meaningful
  clinical audit trails~\cite{selvaraju2017,chattopadhyay2018}.
\end{description}

\noindent These failure modes are not historical artefacts of early deep learning
research. A survey of published systems from 2024--2026 confirms that all three
remain unresolved: state-of-the-art classification approaches based on EfficientNet,
ConvNeXt, and hybrid CNN architectures achieve high accuracy on pre-cropped,
individually segmented cell patches with binary
labels~\cite{mujahid2024,mmileng2025,oladimeji2026,gaouar2025}, but no recent system
operates on whole blood smear images, resolves annotation-incomplete training, or
provides per-cell spatial explainability on the full slide. Concurrently,
Issah \etal~\cite{issah2026} demonstrate that the latest YOLO-based whole-slide
detectors still suffer performance drops exceeding 80\% under cross-dataset
evaluation, confirming that annotation sensitivity~(P1) and generalisation failure
remain active barriers to clinical deployment. This conclusion is consistent with
the recent systematic review by Sukumarran \etal~\cite{sukumarran2024review}, which
identifies binary classification, weak cross-dataset validation, and limited
multi-stage/species classification as persistent gaps in automated malaria
diagnosis from microscopic blood smears.

We present \malariai{}, a two-stage decoupled framework designed to address all three
failure modes. The key insight is that cell
\emph{localisation} and cell \emph{classification} are biologically
distinct sub-problems that benefit from different inductive biases.
By separating them into sequential, independently optimized stages,
\malariai{} avoids the annotation dependency of end-to-end detection
(P1), bypasses the NMS step (P2), and enables per-cell
spatial explainability to be applied after classification (P3).

\medskip
\noindent This paper makes four concrete contributions:

\begin{enumerate}[leftmargin=*, label=\textbf{C\arabic*.}]
  \item \textbf{Label-Resilient Cell Segmentation (C1).}
    An annotation-agnostic Stage~1 based on distance-transform guided
    watershed recovers 75.95\% of ground-truth cells by centroid localisation
    (66.88\% at strict IoU~$\geq 0.5$) on the 120-image BBBC041 test set
    \emph{without consulting any annotation}, structurally eliminating
    Problem~P1 from the pipeline.

  \item \textbf{Density-Invariant Instance Separation (C2).}
    Distance-transform watershed operates at the morphological level
    and does not apply NMS. Touching cells in dense smear regions are
    separated at the instance level by the distance-transform topology,
    bypassing Problem~P2.

  \item \textbf{Integrated End-to-End Spatial Explainability (C3).}
    Grad-CAM\texttt{++} heatmaps are generated for every cell detected
    in a full blood smear image, producing a two-view output: a 160-pixel
    crop overlay and a full-image heatmap localising all parasite
    activations simultaneously. To our knowledge, this is the first
    whole-slide malaria detection system to integrate per-cell spatial
    explainability as a first-class output.

  \item \textbf{Cross-Dataset Generalisation Assessment (C4).}
    The complete pipeline is evaluated zero-shot on MP-IDB (209~images,
    17~patients, four \textit{Plasmodium} species) without any
    hyperparameter re-tuning, providing an honest out-of-distribution
    characterisation that is rare in the malaria detection literature and
    directly quantifies the gap between source-domain performance and
    real-world deployment conditions.
\end{enumerate}

\medskip
\noindent Individually, distance-transform watershed, EfficientNet-B0,
Focal Loss, and Grad-CAM\texttt{++} are each established techniques; the
engineering contribution of \malariai{} is their decoupled integration
(C1--C3). Separately, and on its own footing, Section~\ref{ssec:stage1_alt_metrics}
introduces four complementary localisation metrics (M1--M4) that
disentangle true missed detections from boundary mis-alignment
artefacts specific to organic-shaped segmentation - a measurement
contribution applicable to any watershed-based or otherwise
non-rectangular detector evaluated against box-based ground truth, not
just to \malariai{} itself.

\medskip
\noindent Table~\ref{tab:related_summary} positions \malariai{} against
the most closely related prior systems across the three failure modes
above; no prior system resolves all three jointly.

\begin{table}[!htbp]
\centering
\small
\caption{Positioning of \malariai{} against prior work.
  \textbf{P1} = annotation-agnostic detection;
  \textbf{P2} = dense overlap handling without NMS;
  \textbf{P3} = integrated per-cell spatial XAI on whole-slide images.
  \checkmark~= addressed; $\sim$~= partial; $\times$~= not addressed.}
\label{tab:related_summary}
\renewcommand{\arraystretch}{1.2}
\begin{tabular}{lccc}
\toprule
\textbf{System} & \textbf{P1} & \textbf{P2} & \textbf{P3} \\
\midrule
Singh \etal~\cite{singh2025}              & $\times$ & $\times$ & $\times$ \\
Loh \etal~\cite{loh2021}                  & $\times$ & $\times$ & $\times$ \\
Optimized YOLOv4~\cite{yolov4malaria2024} & $\times$ & $\times$ & $\times$ \\
Delgado-Ortet \etal~\cite{delgado2020}    & $\times$ & $\times$ & $\times$ \\
Pandiaraj \etal~\cite{pandiaraj2024}       & $\times$ & $\sim$   & $\times$ \\
XAI ensemble~\cite{xai_ensemble2025}      & $\times$ & $\times$ & $\sim$   \\
CellSAM~\cite{marks2025cellsam}           & $\times$ & $\checkmark$ & $\times$ \\
Disco~\cite{sun2026disco}                 & $\times$ & $\checkmark$ & $\times$ \\
\midrule
\textbf{\malariai{} (ours)}               & \checkmark & \checkmark & \checkmark \\
\bottomrule
\end{tabular}
\end{table}

\medskip
\noindent\textbf{Paper organisation.}
Section~\ref{sec:related} reviews the related literature across five thematic
areas. Section~\ref{sec:method} describes the dataset, system overview, and
evaluation metrics. Section~\ref{sec:experiments} presents quantitative and
qualitative results, including a zero-shot cross-dataset evaluation on the
held-out MP-IDB benchmark. Section~\ref{sec:discussion} discusses implications and
limitations. Section~\ref{sec:conclusion} concludes with directions for future
work.

%% ============================================================================
%% SECTION 2 -- RELATED WORK
%% ============================================================================
\section{Related Work}
\label{sec:related}

We organize the literature into five thematic areas: (i)~classical and
feature-based approaches, (ii)~CNN-based classification on pre-isolated
cells, (iii)~end-to-end object detection and instance segmentation,
(iv)~two-stage decoupled pipelines, and (v)~explainable AI for clinical
malaria diagnosis. We conclude with a cross-cutting analysis of open
problems.

\subsection{Classical and Feature-Based Approaches}
\label{ssec:classical}

Early automated malaria diagnosis relied on carefully engineered image
processing pipelines applied to Giemsa- or May--Gr\"{u}nwald--Giemsa
(MGG)-stained blood smears. The dominant approach combined global
thresholding (\eg, Otsu's method~\cite{otsu1979}), morphological
operations such as erosion and dilation, and hand-crafted feature
descriptors (\eg, shape moments, color histograms, Gabor texture
features) fed into classical classifiers such as Support Vector Machines
(SVM) or Random Forests~\cite{poostchi2018}.

These pipelines established a fundamental insight that malaria detection
is a two-phase problem: cells must first be localized before their internal
morphology can be assessed. However, they suffered from two compounding
weaknesses. Segmentation quality degraded sharply when cells overlapped,
and hand-crafted features failed to generalize across staining protocols,
microscope configurations, and acquisition sites~\cite{poostchi2018,delgado2020}.
Nonetheless, classical watershed segmentation based on the distance
transform~\cite{beucher1992} remains competitive for the cell
\emph{localisation} sub-problem: it requires no labeled training data,
its convergence is deterministic, and it is naturally suited to the circular
morphology of red blood cells. This motivates our adoption of
distance-transform watershed as Stage~1.

Most closely related to our Stage~1 design, Angkoso
\etal~\cite{angkoso2026array} benchmark five classical segmentation
techniques -- Watershed, Fuzzy C-Means, K-means, Gaussian Mixture
Models, and Entropy Filtering -- within a color-cascading preprocessing
framework (RGB normalization, gamma correction, noise reduction,
exposure compensation, edge enhancement) on 574 microscopy images
spanning four \textit{Plasmodium} species. Fuzzy C-Means is reported as
the most reliable method overall (98.26\% accuracy, 97.91\% sensitivity,
98.61\% specificity), with Watershed competitive on well-contrasted
images but comparatively less consistent on faint or overlapping
structures. This result is complementary rather than overlapping with
our contribution: Angkoso \etal{} evaluate segmentation quality as a
pixel-wise classification task (parasite vs.\ non-parasite pixels) on a
private clinical dataset, with no cell-level detection output, no
species or lifecycle-stage classification, and no deep learning
component -- their own stated future work explicitly identifies
integrating a deep learning architecture such as U-Net and adding stage
classification as open directions. \malariai{} addresses precisely
this gap: Stage~1's watershed output is annotation-free cell-level
\emph{detection} (bounding boxes, not pixel masks) feeding a downstream
7-class deep learning classifier (Stage~2), validated on public
benchmarks (BBBC041, MP-IDB) with cross-dataset generalisation and
per-cell explainability, none of which Angkoso \etal{} address. The two
studies are best read as adjacent: Angkoso \etal{} establish that
Watershed is a reasonable but not uniquely dominant classical
segmentation choice among non-learned alternatives, which is consistent
with our own decision (Section~\ref{ssec:unet_baseline}) to benchmark
Stage~1's watershed against a learned alternative rather than assume
its superiority.

\subsection{CNN-Based Classification of Pre-Isolated Cells}
\label{ssec:cnn_class}

The release of the NIH malaria cell image dataset~\cite{rajaraman2019},
comprising individually cropped and labeled cell patches, catalyzed CNN
classification studies. Rajaraman \etal~\cite{rajaraman2019} evaluated
fine-tuned architectures (VGG-16, ResNet-50, Xception) and found that
transfer learning from natural image corpora substantially outperformed
training from scratch.

More recently, Singh \etal~\cite{singh2025} propose an optimized hybrid
framework combining a 12-layer CNN with EfficientNet-B7 through a parallel
feature-fusion head. Otsu-based preprocessing of the RGB images enhances
parasite-relevant regions before classification, yielding 97.96\% accuracy
on their 43,400-image dataset. Segmentation quality on a manually annotated
100-image subset achieves a Dice coefficient of 0.848 and Jaccard Index
of 0.738. The approach is conceptually related to our Stage~2 design, but
it operates exclusively on pre-cropped, individually isolated cell patches.
The cell isolation problem - finding cells in a dense, unannotated whole
slide - is absent. The authors explicitly acknowledge that
``interpretability tools need to be refined to become clinically
applicable'' and identify real-time deployment as future work~\cite{singh2025}.

This family of work collectively avoids the hard sub-problem of whole-slide
cell detection. Reported accuracy metrics measure only classification on
pre-isolated patches and cannot be compared to detection system benchmarks.

\subsection{End-to-End Object Detection and Instance Segmentation}
\label{ssec:e2e}

The Faster R-CNN architecture~\cite{ren2015} - combining a Region Proposal
Network (RPN) with a two-stage classification head over a
ResNet-50~\cite{he2016resnet} Feature Pyramid Network (FPN)~\cite{lin2017fpn}
backbone - was among the first deep learning systems applied directly to
the whole-slide malaria detection problem. Hung \etal~\cite{hung2018}
demonstrated high accuracy using a cascaded Faster R-CNN with an AlexNet
backbone, establishing the feasibility of end-to-end detection. Subsequent
work has explored lighter architectures: optimized YOLOv4 models with
backbone replacement and layer pruning achieve strong mAP on thin
smear images~\cite{yolov4malaria2024}. Rather than only citing this
trajectory toward faster single-stage detectors, Section~\ref{ssec:baseline_b}
additionally trains and evaluates YOLOv8s~\cite{yolov8} directly on our
data as Baseline B, alongside Faster R-CNN Baseline A.

Instance segmentation - producing a pixel-level mask per detected cell
- was explored by Loh \etal~\cite{loh2021} using Mask R-CNN~\cite{he2017maskrcnn}.
Their model achieves inference speeds 15$\times$ faster than manual counting.
These results represent the closest prior art to our Baseline A evaluation.
Transformer detectors provide a complementary NMS-free direction:
DN-DETR~\cite{li2022dndetr} accelerates DETR training by denoising
ground-truth queries while retaining set-based prediction and Hungarian
matching. However, such detectors still require complete bounding-box
supervision during training. In a sparsely annotated smear dataset, the
closed-world label problem therefore remains even if NMS is removed.
In malaria specifically, Guemas \etal~\cite{guemas2024rtdetr} evaluate
a real-time detection transformer (RT-DETR) for patient-level recognition
of four \textit{Plasmodium} species from thin blood smears, using a large
manually labelled corpus across multiple hospitals. This work confirms that
transformer-style detectors are entering malaria microscopy, but its focus
is species recognition from labelled thin-smear data rather than
annotation-agnostic universal cell recovery, dense-cell instance separation,
or per-cell spatial explainability.

However, end-to-end detectors exhibit two systematic failure modes in the
malaria domain. First, all such models operate under the closed-world
assumption: any region without a ground-truth bounding box is treated as
background during training. In NIH BBBC041, annotation is incomplete; a
detector trained on this data suppresses unannotated cells, producing recall
figures that reflect annotation density rather than true cell recovery rate.
Loh \etal~\cite{loh2021} acknowledge this, proposing semi-automated
labeling to reduce annotation bias. This limitation remains an active
concern: Issah \etal~\cite{issah2026} demonstrate that state-of-the-art
YOLOv12 architectures trained on curated malaria datasets suffer
performance drops exceeding 80\% under cross-dataset evaluation,
confirming that end-to-end detectors remain sensitive to the annotation
characteristics of their training sets when deployed in new settings.
The annotation incompleteness problem~(P1) is not confined to the malaria
domain: Bai \etal~\cite{bai2023} document that the widely-used DeepLesion
benchmark has a missing annotation rate of approximately 50\%, and demonstrate
that training end-to-end detectors directly on such data yields suboptimal
recall~\cite{bai2023}. Their ET-ULD framework addresses this through iterative
pseudo-label mining, but still requires a reliably annotated initial subset.
\malariai{}'s annotation-agnostic watershed requires no labeled data at
the detection stage, structurally eliminating this dependency.

Second, Non-Maximum Suppression (NMS) calibrated for non-overlapping objects
fails in dense smear regions where genuine cells share high IoU. Rather than
solving this, prior work filters out dense images before
evaluation~\cite{loh2021}, sacrificing exactly the images where automated
diagnosis is most needed. This failure mode has been independently confirmed
at the highest-impact venues: Marks \etal~\cite{marks2025cellsam} explicitly
state that R-CNN-family detectors ``rely on non-maximum suppression (NMS),
leading to missed instances in crowded scenes with irregular morphologies or
high degrees of overlap'' and present CellSAM -- a SAM-based segmentation
framework built on the promptable Segment Anything foundation
model~\cite{kirillov2023sam} -- as a workaround requiring pixel-level mask
annotations that are unavailable in the malaria setting. Formally,
Sun \etal~\cite{sun2026disco}
conduct the first systematic analysis of cell adjacency graph topology across
four histopathology benchmarks, demonstrating that real-world cellular
arrangements are fundamentally non-bipartite -- characterized by odd-length
cycles that make NMS-based instance disambiguation inherently infeasible.
Their measured conflict-node ratio reaches 30.49\% in the GBC-FS 2025
gallbladder dataset, providing a rigorous graph-theoretic foundation for
our empirical finding that 58\% of BBBC041 images contain ground-truth
cell pairs with $\text{IoU} > 0.3$.

\subsection{Two-Stage Decoupled Pipelines}
\label{ssec:twostage}

The closest architectural precedents to \malariai{} separate cell
segmentation from cell classification into sequential stages. Encoder-decoder
architectures --- pioneered by U-Net~\cite{ronneberger2015unet} for biomedical
image segmentation and extended by SegNet~\cite{badrinarayanan2017} and
UNet++~\cite{zhou2018unetpp} --- provide the segmentation backbone underpinning
most such pipelines. Rather than adopt one of these supervised segmentation
networks as Stage~1, \malariai{} instead uses an annotation-free watershed
algorithm; Section~\ref{ssec:unet_baseline} reports a direct, same-test-set
quantitative comparison against a trained U-Net baseline to characterize
this design choice empirically rather than by assumption. More recently, CellViT~\cite{horst2024cellvit} demonstrates
that Vision Transformer encoders can jointly address cell segmentation and
classification in histopathology, though at substantially higher computational
cost than a decoupled approach.

\textbf{Delgado-Ortet \etal}~\cite{delgado2020} present a three-stage
pipeline comprising a SegNet~\cite{badrinarayanan2017}-based encoder-decoder segmentation network,
a crop-and-mask stage, and a 13-layer CNN classifier, achieving
93.72\% global segmentation accuracy and 87.04\% malaria detection
specificity on MGG-stained smears. The work demonstrates decoupled pipeline
viability but suffers from three failures: the segmentation network
misclassifies leukocytes and platelets, inducing cascading classification
errors; the classifier overfits from 95.0\% validation to 75.39\% test
accuracy due to a five-patient training set; and class imbalance is
unaddressed. Their stated future directions encompass annotating additional
cell classes, generating larger datasets, and mixing data across
institutions; taken together, these align closely with the improvements
\malariai{} implements.

More recently, \textbf{Pandiaraj \etal}~\cite{pandiaraj2024} combine a
Trans-MobileUNet++ segmentation network with an Adaptive and
Atrous Convolution-based Recurrent MobileNetV2 detection backbone, reporting
94.88\% accuracy on Kaggle-hosted pre-cropped cell image datasets. However,
the framework performs binary classification only (parasitized vs.\ uninfected),
does not operate on whole blood smear images, and reports no mAP metric at
any IoU threshold, precluding quantitative comparison with standardized
detection benchmarks. Furthermore, their metaheuristic optimization introduces
severe reproducibility concerns: sensitivity to random initialization and the
absence of convergence guarantees makes independent replication
impractical~\cite{pandiaraj2024}.

Consequently, while the two-stage literature suggests that architectural
decoupling is the correct design direction, no existing system jointly
addresses annotation resilience, dense-region recall, multi-class imbalance,
and spatial explainability.

\subsection{Explainable AI for Clinical Malaria Diagnosis}
\label{ssec:xai}

Physician trust in AI systems depends on the availability of spatial
evidence for model predictions~\cite{xai_ensemble2025}. Gradient-weighted
Class Activation Mapping (Grad-CAM)~\cite{selvaraju2017} and its
extension Grad-CAM\texttt{++}~\cite{chattopadhyay2018} produce spatial
heatmaps by weighting convolutional feature maps by class-specific gradient
signals. In the malaria context, these maps should ideally localize the
parasite body within the infected cell.

Several recent works demonstrate meaningful Grad-CAM outputs on malaria
cell images. Transformer-based models with compact convolutional heads
produce interpretable attention maps for parasite
detection~\cite{xai2022transformer}. Ensemble frameworks with XAI
integration have been benchmarked against clinical diagnostic
criteria~\cite{xai_ensemble2025}. However, \emph{all} existing XAI
applications in the malaria literature operate on pre-cropped single-cell
patches. No published system integrates spatial explainability into a
pipeline that performs whole-slide detection, cell segmentation,
classification, and explanation end-to-end. \malariai{} closes this gap:
Grad-CAM\texttt{++} is applied within the Stage~2 classifier to crops
produced by Stage~1, making every cell in a full blood smear image
explainable without requiring pre-isolation. Detection-specific XAI methods
such as D-RISE~\cite{petsiuk2021drise} show that object-detector predictions
can be explained with saliency maps, but their black-box masking strategy
requires many perturbed forward passes per target detection. This makes them
useful for detector auditing but less suited to routine whole-slide malaria
screening, where dozens or hundreds of cell instances may require explanation
in a single smear. The broader importance of whole-slide spatial analysis over
patch-level processing is affirmed by
recent computational pathology infrastructure: Zheng \etal~\cite{zheng2026lazyslide}
introduce LazySlide (Nature Methods, 2026), identifying fragmented
patch-based workflows as a primary barrier to clinical adoption in digital
pathology, and establishing whole-slide cell segmentation and zero-shot
classification as the target capability. \malariai{} achieves this in the
malaria domain through its integrated detection, classification, and
Grad-CAM\texttt{++} pipeline on full 1600$\times$1200 blood smear images.

\subsection{Cross-Cutting Open Problems and Positioning}
\label{ssec:gaps}

Table~\ref{tab:related_summary} (Section~\ref{sec:intro}) summarizes how
the most closely related prior works address the three open problems.
In addition, the cross-dataset setting addressed in Contribution~C4 is a
known barrier across medical image analysis: Guan and Liu~\cite{guan2022domain}
survey domain adaptation methods and identify domain shift between source and
target medical image distributions as a central obstacle to robust deployment.
This directly contextualizes our zero-shot BBBC041-to-MP-IDB experiment:
\malariai{} does not claim domain adaptation is solved, but explicitly measures
the shift and exposes where Stage~1 requires dataset-adaptive parameters.
Recent malaria dataset releases reinforce the importance of this direction:
Nakasi \etal~\cite{nakasi2025dataset} publish a Uganda mobile-microscopy
benchmark containing annotated thick and thin blood smear images, while also
noting thin-smear limitations in species and life-cycle diversity. Such datasets
are natural future test beds for \malariai{}'s cross-dataset evaluation, but they
do not remove the need to report source-to-target performance honestly.

%% ============================================================================
%% SECTION 3 -- SYSTEM OVERVIEW AND METHODOLOGY
%% ============================================================================
\section{System Overview and Methodology}
\label{sec:method}

\subsection{Dataset Description}
\label{ssec:dataset_group}

\subsubsection{Primary Dataset: BBBC041}
\label{ssec:dataset}

\paragraph{Dataset.}
All experiments use the NIH BBBC041 benchmark~\cite{bbbc041}, a publicly
available collection of Giemsa-stained thin \textit{Plasmodium falciparum}
blood smear images acquired at $100\times$ oil-immersion magnification.
The dataset comprises 1,208 training images and 120 held-out test images,
each at $1{,}600 \times 1{,}200$ pixels (RGB). Raw annotations total
80,113 bounding boxes in training and 5,922 in testing across seven
categories. Following standard practice, \textit{difficult} annotations
(441 training; 5 test) are excluded from all evaluations, yielding 79,672
and 5,917 valid boxes respectively.

\begin{figure}[!htbp]
  \centering
  \includegraphics[width=\columnwidth]{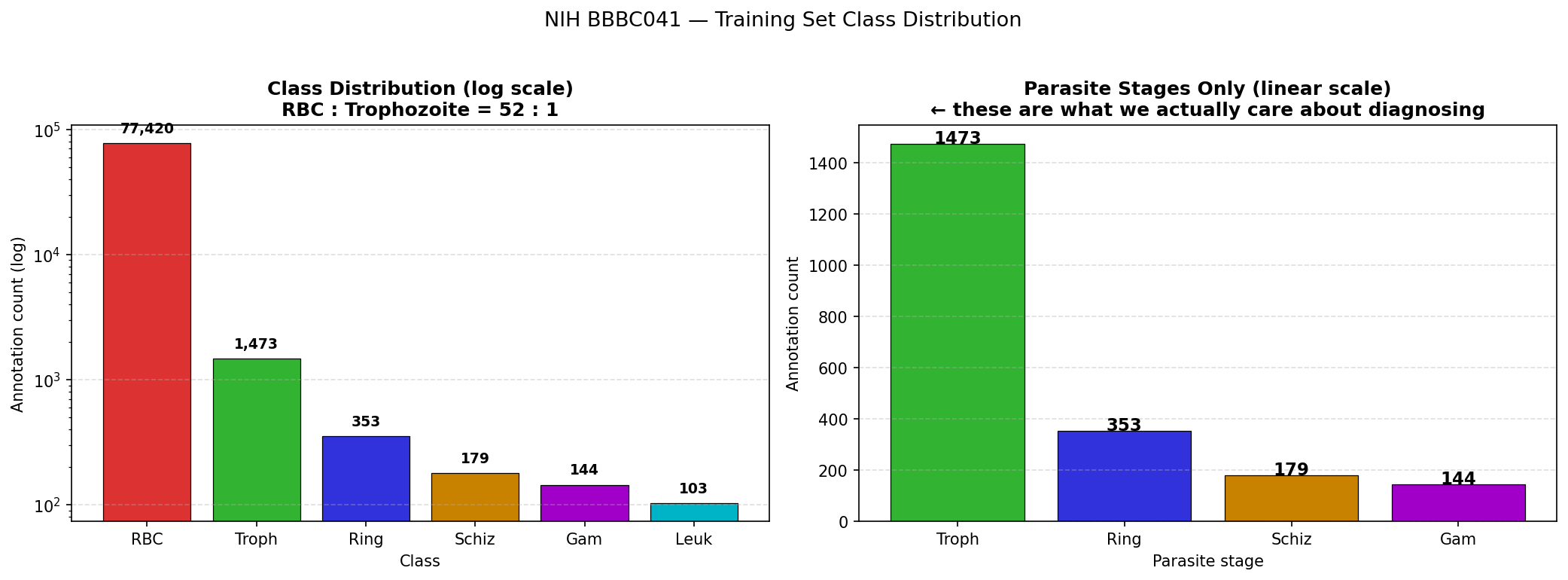}
  \caption{Class distribution of NIH BBBC041 training annotations.
    Left: log-scale view dominated by the red blood cell class (77,420
    instances). Right: linear-scale view of parasitic stages only, showing
    the secondary imbalance between trophozoites (1,473) and the rarest
    stages (gametocyte: 144). The 537:1 RBC-to-gametocyte ratio motivates
    per-class Focal Loss weighting.}
  \label{fig:class_dist}
\end{figure}

\paragraph{Class distribution and imbalance.}
Table~\ref{tab:class_dist} and Figure~\ref{fig:class_dist} report per-class
annotation counts. The distribution is severely imbalanced: red blood cells
constitute 97.2\% of training annotations, while gametocytes account for
only 0.18\% - a ratio of 537:1. A model predicting \textit{red blood cell}
for every instance achieves 97.2\% accuracy while detecting zero parasites.
This motivates Focal Loss~\cite{lin2017focal} with per-class inverse
frequency weights in Stage~2 (Section~\ref{ssec:stage2}).

\begin{table}[!htbp]
\centering
\small
\caption{Per-class annotation counts in NIH BBBC041
  (\textit{difficult} annotations excluded).}
\label{tab:class_dist}
\renewcommand{\arraystretch}{1.15}
\begin{tabular}{lrrl}
\toprule
\textbf{Class} & \textbf{Train} & \textbf{Test} & \textbf{\% of Train} \\
\midrule
Red blood cell  & 77{,}420 & 5{,}614 & 97.18\% \\
Trophozoite     &  1{,}473 &    111  &  1.85\% \\
Ring            &    353   &    169  &  0.44\% \\
Schizont        &    179   &     11  &  0.22\% \\
Gametocyte      &    144   &     12  &  0.18\% \\
Leukocyte       &    103   &      0  &  0.13\% \\
\midrule
\textbf{Total}  & 79{,}672 &  5{,}917 & \\
\bottomrule
\end{tabular}
\end{table}

\begin{figure}[!htbp]
  \centering
  \includegraphics[width=\columnwidth]{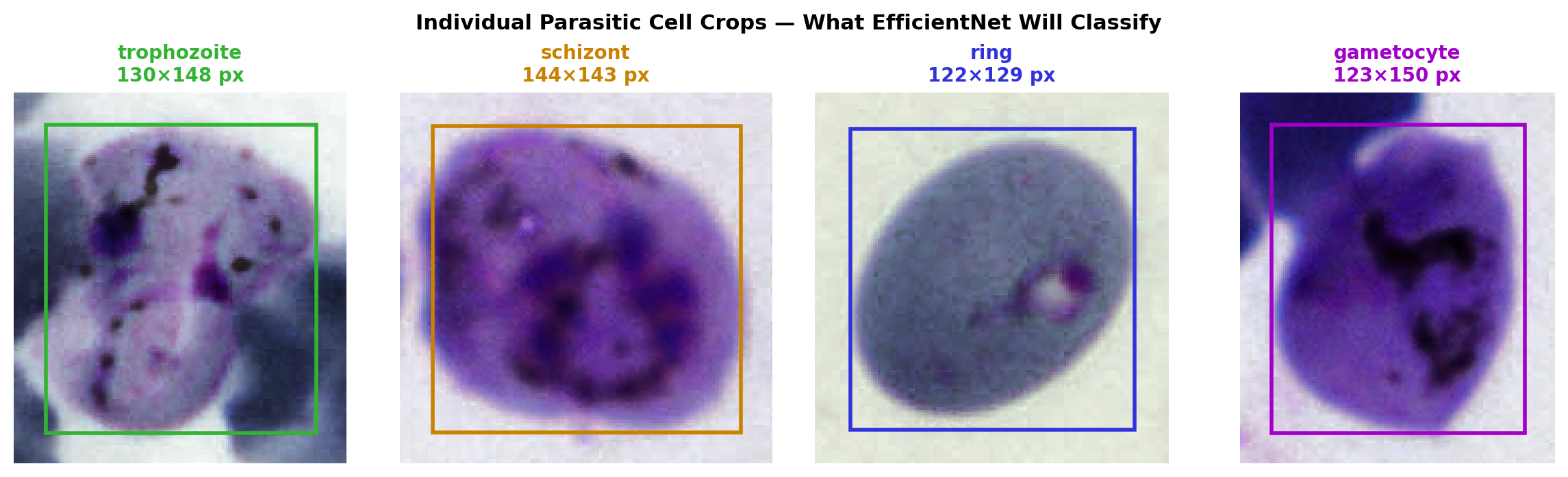}
  \caption{Representative ground-truth crops of each parasitic stage in
    NIH BBBC041 at native resolution. Ring-stage infections are visually
    the most similar to healthy RBCs, predicting lower per-class accuracy
    for that class. Parasitic stages are systematically larger than healthy
    erythrocytes (trophozoite median area: 17,272\,px$^2$; RBC: 11,544\,px$^2$),
    consistent with intraerythrocytic swelling during parasite development.}
  \label{fig:parasite_crops}
\end{figure}

\paragraph{Parasitic stage morphology.}
Figure~\ref{fig:parasite_crops} shows representative crops for each
parasitic stage. Ring-stage infections present as pale discs with only a
faint peripheral chromatin outline, closely resembling uninfected RBCs.
Trophozoites exhibit irregular chromatin mass; schizonts show clear
internal segmentation (merozoites); gametocytes appear as large, densely
stained elongated bodies. This morphological hierarchy directly predicts
which classes the Stage~2 classifier will find most challenging.

\begin{figure}[!htbp]
  \centering
  \includegraphics[width=\columnwidth]{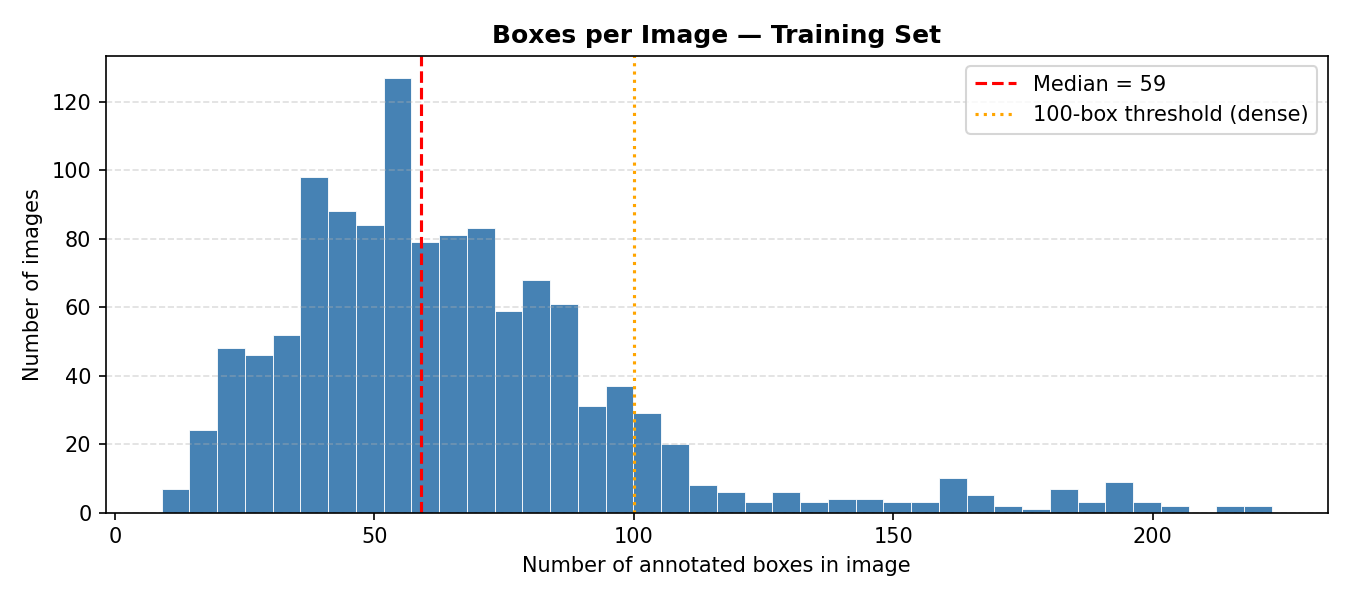}
  \caption{Distribution of annotated bounding boxes per image in the
    BBBC041 training set (1,208 images). The dashed red line marks the
    median (59 boxes); the dotted orange line at 100 boxes identifies
    images we classify as \emph{dense smears}. 11\% of images exceed
    this threshold. 58\% of a 200-image sample contain at least one
    overlapping cell pair ($\text{IoU}>0.3$), confirming the severity of
    the NMS failure mode.}
  \label{fig:density}
\end{figure}

\paragraph{Annotation density and cell overlap.}
Figure~\ref{fig:density} shows the distribution of annotated boxes per
image. Density ranges from 9 to 223 boxes, with a median of 59 and a
95th percentile of 137; 11\% of images exceed 100 cells. To quantify
Problem~P2, we computed pairwise IoU between all ground-truth boxes within
each image over a 200-image sample. \textbf{58\%} of images contain at
least one ground-truth pair with $\text{IoU} > 0.3$, confirming that cell
overlap is a structural feature of the data. This empirical finding
directly motivates distance-transform watershed as Stage~1.

\paragraph{Train/validation/test protocol.}
The 1,208 training images are partitioned 80/20 at the image level into
training (966 images) and validation (242 images) splits. No image
contributes boxes to both sets. The random seed is fixed at 42 for
reproducibility. All hyperparameter decisions use the validation split
exclusively. The 120 test images are held out for Stage~1 evaluation only.

%% --------------------------
\subsubsection{Cross-Dataset Validation: MP-IDB}
\label{ssec:mpidb}

To assess generalisation beyond the NIH BBBC041 benchmark, we adopt the
\textit{Malaria Parasite Image Database} (MP-IDB)~\cite{loddo2019mpidb}
as a held-out cross-dataset evaluation set. MP-IDB comprises 209~images
from 17~distinct patient samples, acquired at 2{,}592$\times$1{,}944
pixels under Giemsa thin-smear staining - the same staining protocol as
BBBC041, but a different institution, magnification calibration, and
patient cohort. Annotations were released in Supervisely bitmap-mask
format; we decode each mask via base64~$\rightarrow$~zlib~$\rightarrow$~PNG
and derive tight bounding boxes from the mask origin and dimensions,
yielding 1{,}407 annotated infected cells (script: \texttt{data/prepare\_mpidb.py}).

\paragraph{Taxonomy mismatch.}
MP-IDB labels cells at the \textit{species} level
(\textit{P.\ falciparum}, \textit{P.\ vivax}, \textit{P.\ malariae},
\textit{P.\ ovale}), whereas BBBC041 labels cells at the \textit{lifecycle
stage} level (ring, trophozoite, schizont, gametocyte). These taxonomies
are orthogonal: species identity does not determine lifecycle stage.
Accordingly, MP-IDB evaluation is \textit{binary}: any annotated object
is an infected cell (positive), and the pipeline's task is to detect it
regardless of species or stage.

\begin{figure}[!htbp]
  \centering
  \includegraphics[width=\columnwidth]{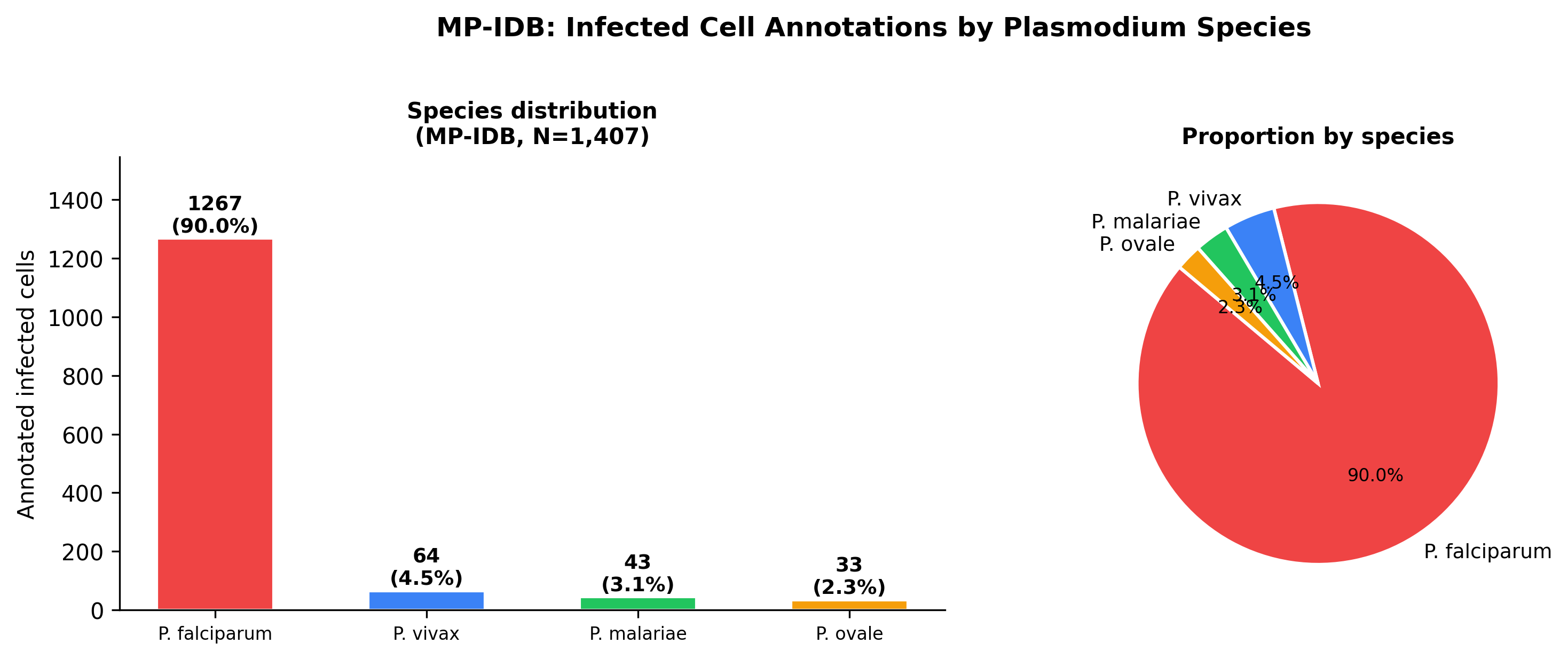}
  \caption{MP-IDB species distribution. Left: annotation counts per
    \textit{Plasmodium} species across 209 images.
    \textit{P.\ falciparum} comprises 90.0\% of all annotated infected
    cells (1{,}267 of 1{,}407), consistent with its epidemiological
    prevalence. Right: proportional breakdown. The severe species imbalance
    mirrors the stage imbalance in BBBC041 and motivates binary (infected
    vs.\ uninfected) cross-dataset evaluation.}
  \label{fig:mpidb_class_dist}
\end{figure}

\begin{figure}[!htbp]
  \centering
  \includegraphics[width=\columnwidth]{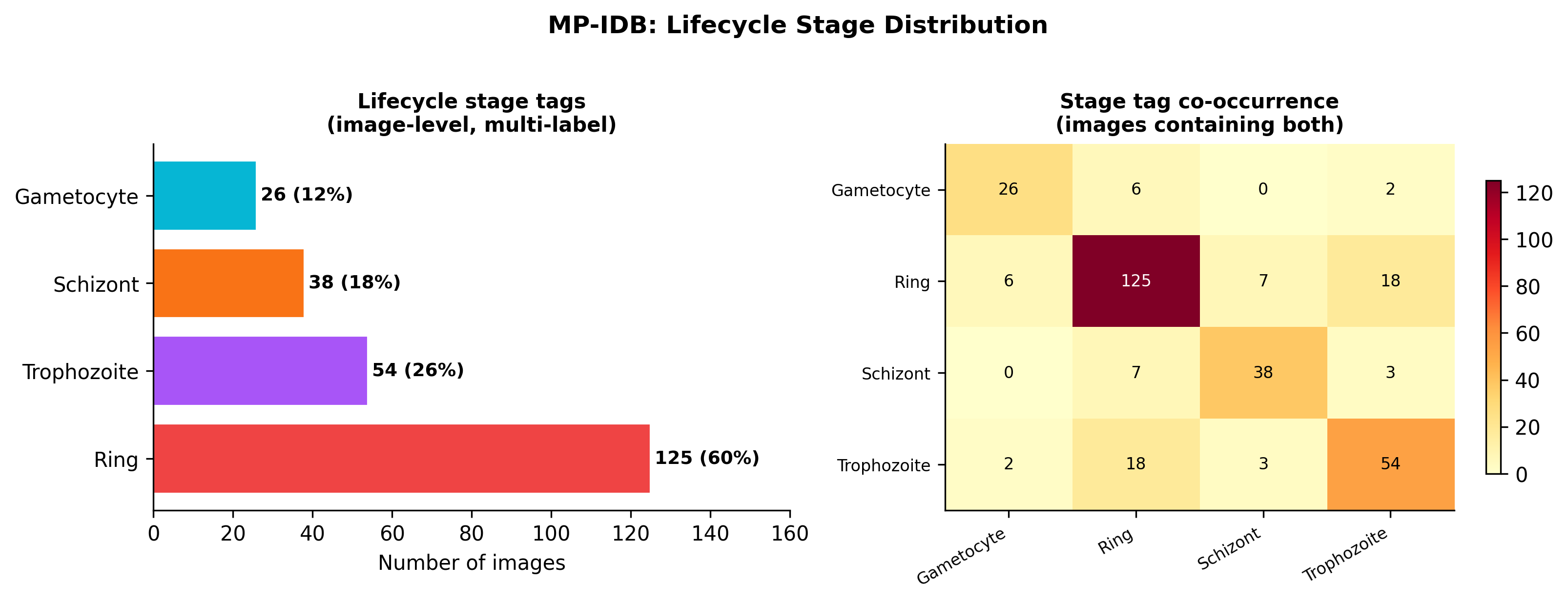}
  \caption{MP-IDB lifecycle stage distribution. Left: number of images
    containing each stage tag (multi-label; one image may carry multiple
    tags). Ring-stage infections are the most common (60\% of images).
    Right: tag co-occurrence matrix - the 18 images tagged with both ring
    and trophozoite indicate mixed-stage smears, reflecting real clinical
    variability. Stage tags are image-level only; individual infected cells
    are not labeled by stage.}
  \label{fig:mpidb_stage_tags}
\end{figure}

\paragraph{Species and stage distribution.}
Figure~\ref{fig:mpidb_class_dist} shows the species breakdown.
\textit{P.\ falciparum} dominates (1{,}267 cells, 90.0\%), reflecting its
prevalence in clinical thin-smear datasets globally. Figure~\ref{fig:mpidb_stage_tags}
reports image-level lifecycle stage tags (multi-label): ring stage appears
in 125 images~(60\%), trophozoite in 54~(26\%), schizont in 38~(18\%),
and gametocyte in 26~(12\%), providing cross-stage diversity despite the
species-only annotation granularity.

\begin{figure}[!htbp]
  \centering
  \includegraphics[width=\columnwidth]{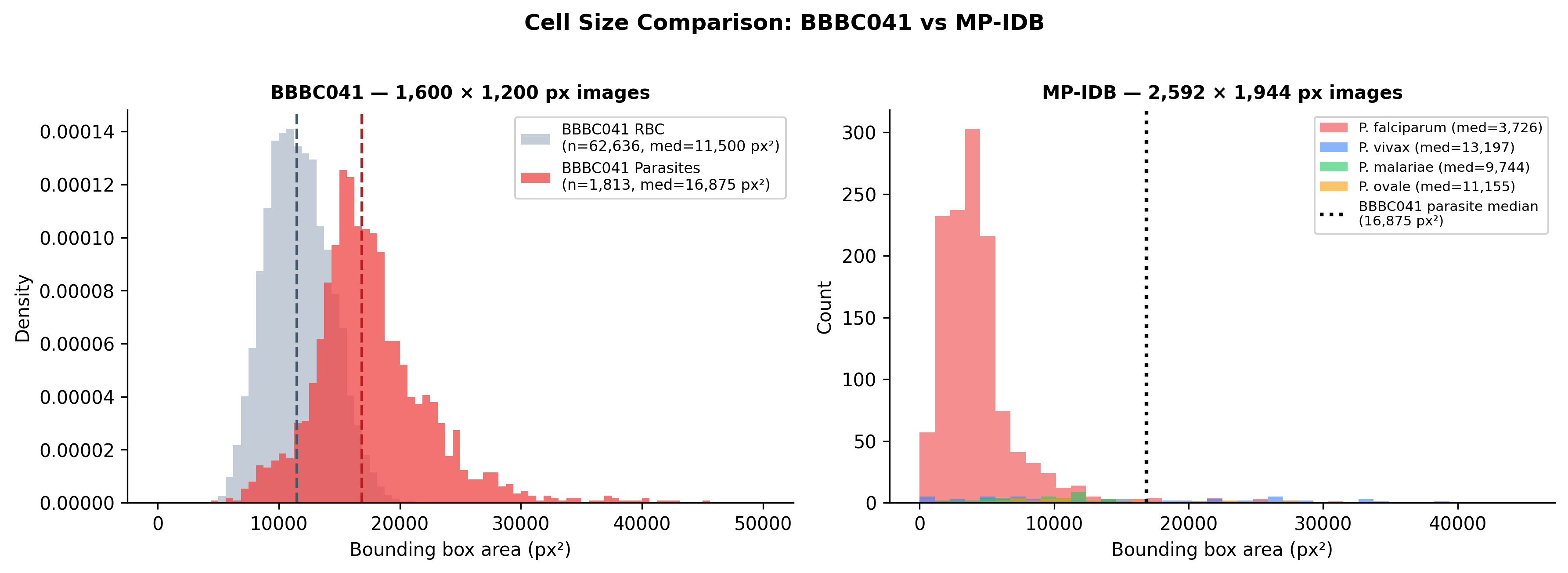}
  \caption{Bounding-box area distributions for BBBC041 (left) and MP-IDB
    (right). Left: BBBC041 RBC and parasite distributions overlap
    substantially; the parasite median (16{,}875\,px$^2$) lies
    well above the RBC median (11{,}500\,px$^2$), reflecting intraerythrocytic
    swelling during parasite development. Right: MP-IDB infected-cell areas
    cluster at far lower values (median 3{,}933\,px$^2$ for
    \textit{P.\ falciparum}), 4.3$\times$ smaller than the BBBC041
    parasite median (dotted line). \textit{P.\ vivax} and \textit{P.\ ovale}
    show larger median areas (13{,}197 and 11{,}155\,px$^2$ respectively),
    consistent with their known tendency to infect larger reticulocytes.
    This scale gap is expected to reduce Stage~1 recall on MP-IDB when
    applied without parameter re-tuning.}
  \label{fig:mpidb_cell_size}
\end{figure}

\begin{figure}[!htbp]
  \centering
  \captionsetup{font=footnotesize}
  \includegraphics[width=\columnwidth]{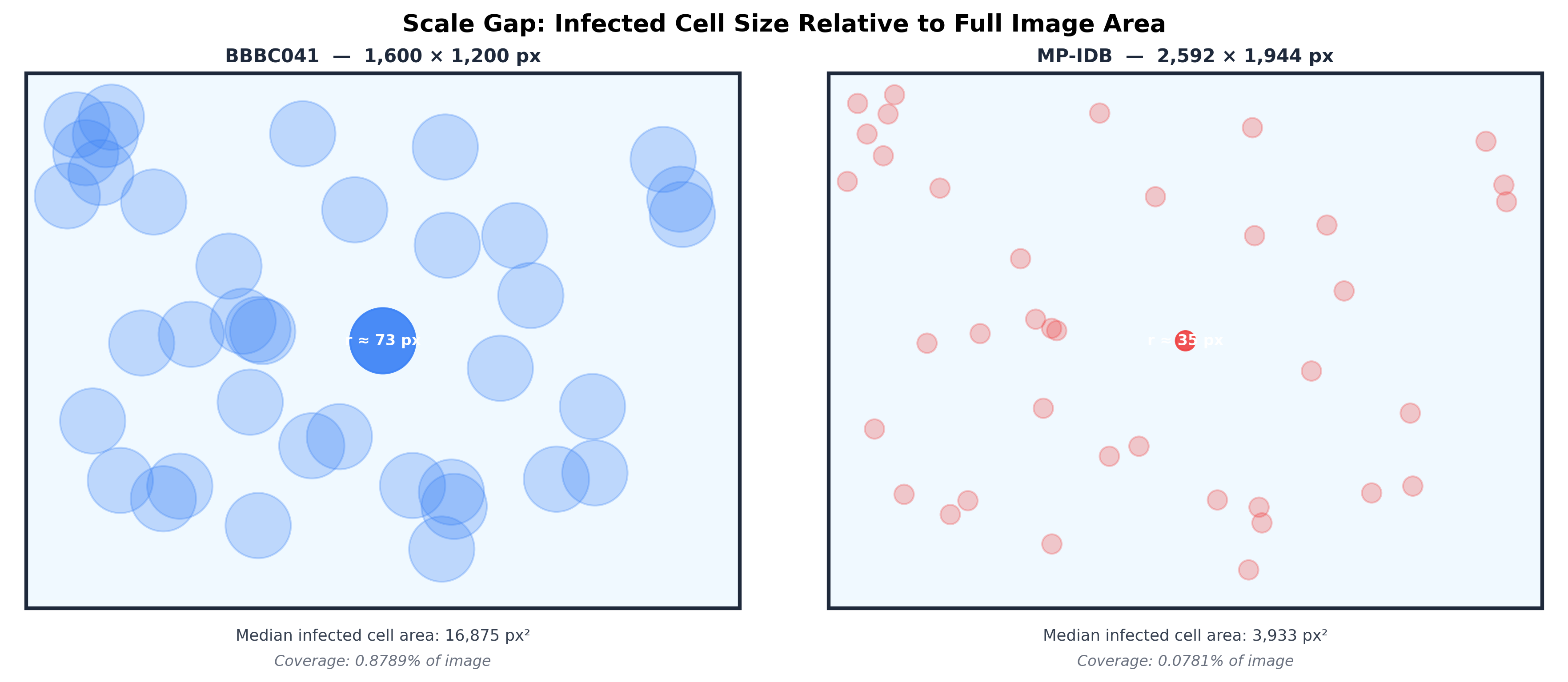}
  \caption{Geometric illustration of the scale gap between BBBC041 (left)
    and MP-IDB (right). Each circle represents an infected cell drawn to
    scale relative to the image frame. In BBBC041 the highlighted cell
    (radius $r \approx 73$\,px) covers 0.88\% of the image; in MP-IDB the
    corresponding cell (radius $r \approx 35$\,px) covers only 0.078\%.
    Watershed parameters tuned on BBBC041 underperform on MP-IDB without
    resolution-aware adaptation.}
  \label{fig:mpidb_scale_gap}
\end{figure}

\paragraph{Resolution and scale gap.}
A critical difference from BBBC041 is the bounding-box area of infected
cells. Figure~\ref{fig:mpidb_cell_size} shows that the median infected-cell
area in MP-IDB is 3{,}933\,px$^2$, compared with 16{,}875\,px$^2$ for
parasitic-stage cells in BBBC041 - a 4.3$\times$ reduction in absolute
size. This is not a biological difference: cells are the same physical
size in both datasets. Rather, it reflects a difference in pixel density
per cell: MP-IDB images contain 2.63$\times$ more pixels overall
(5.04\,Mpx vs.\ 1.92\,Mpx) but fewer magnification-normalized pixels per
cell, so each infected cell covers only 0.078\% of the image frame versus
0.879\% in BBBC041 - an 11.3$\times$ reduction in relative prominence.
Figure~\ref{fig:mpidb_scale_gap} illustrates this geometrically.

This scale gap has a direct consequence for Stage~1: the watershed
hyperparameters \texttt{MIN\_DIST}~$= 12$\,px and \texttt{MIN\_AREA}~$=
150$\,px$^2$ were tuned on BBBC041. For MP-IDB, the resolution-corrected
\texttt{MIN\_DIST} is approximately 20\,px, and the characteristic
infected-cell radius is only 35\,px versus 73\,px in BBBC041. Cross-dataset
evaluation on MP-IDB therefore provides an \emph{honest} assessment of
the pipeline's out-of-distribution behavior without any parameter
re-tuning, and simultaneously motivates the Stage~1 adaptive improvements
described in future work (Section~\ref{ssec:limits}).

Table~\ref{tab:dataset_comparison} summarizes the key properties of both
datasets side by side.

\begin{table*}[tp]
\centering
\small
\caption{Comparison of primary (BBBC041) and cross-validation (MP-IDB) datasets.}
\label{tab:dataset_comparison}
\renewcommand{\arraystretch}{1.1}
\begin{tabular*}{\textwidth}{@{\extracolsep{\fill}}lll@{}}
\toprule
\textbf{Property} & \textbf{BBBC041 (primary)} & \textbf{MP-IDB (cross-val)} \\
\midrule
Source              & NIH / Broad Institute     & Delgado-Ortet et al.\ (2020) \\
Images              & 1{,}328 (train + test)    & 209 (annotated) \\
Resolution          & 1{,}600$\times$1{,}200\,px  & 2{,}592$\times$1{,}944\,px \\
Annotation type     & Bounding boxes            & Bitmap masks $\rightarrow$ boxes \\
Label taxonomy      & Stage-level               & Species-level \\
Infected cells      & 1{,}841                   & 1{,}407 \\
Median cell area    & 16{,}875\,px$^2$          & 3{,}933\,px$^2$ \\
Cell image coverage & 0.879\%                   & 0.078\% \\
Patients            & Not reported              & 17 \\
Staining            & Giemsa thin smear         & Giemsa thin smear \\
\bottomrule
\end{tabular*}
\end{table*}

\begin{figure}[!htbp]
  \centering
  \includegraphics[width=\columnwidth]{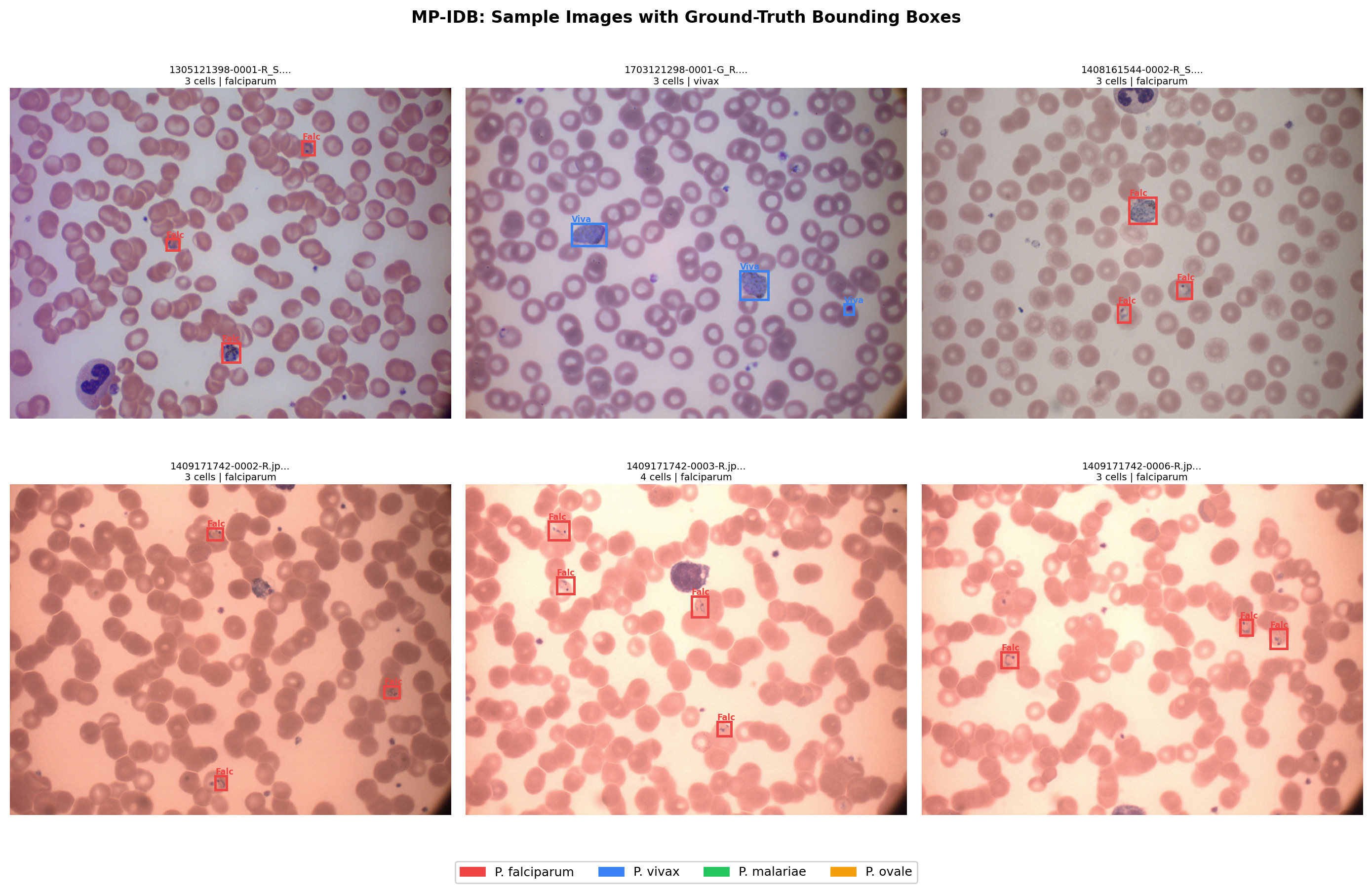}
  \caption{Representative MP-IDB images with ground-truth bounding boxes
    overlaid. Red boxes: \textit{P.\ falciparum}; blue: \textit{P.\ vivax};
    green: \textit{P.\ malariae}; yellow: \textit{P.\ ovale}. Images span
    multiple patients and lifecycle stages. The top row shows standard
    Giemsa staining; the bottom row shows images from a patient with slightly
    different background coloration, illustrating intra-cohort staining
    variability. Infected cells are visually subtle relative to the large
    number of healthy RBCs, confirming the detection challenge.}
  \label{fig:mpidb_samples}
\end{figure}
% \FloatBarrier

%% --------------------------
\subsection{Proposed MalariAI Framework}
\label{ssec:methods_group}

\subsubsection{Overall Pipeline Architecture}
\label{ssec:arch}

\begin{figure}[!htbp]
  \centering
  \includegraphics[width=\linewidth]{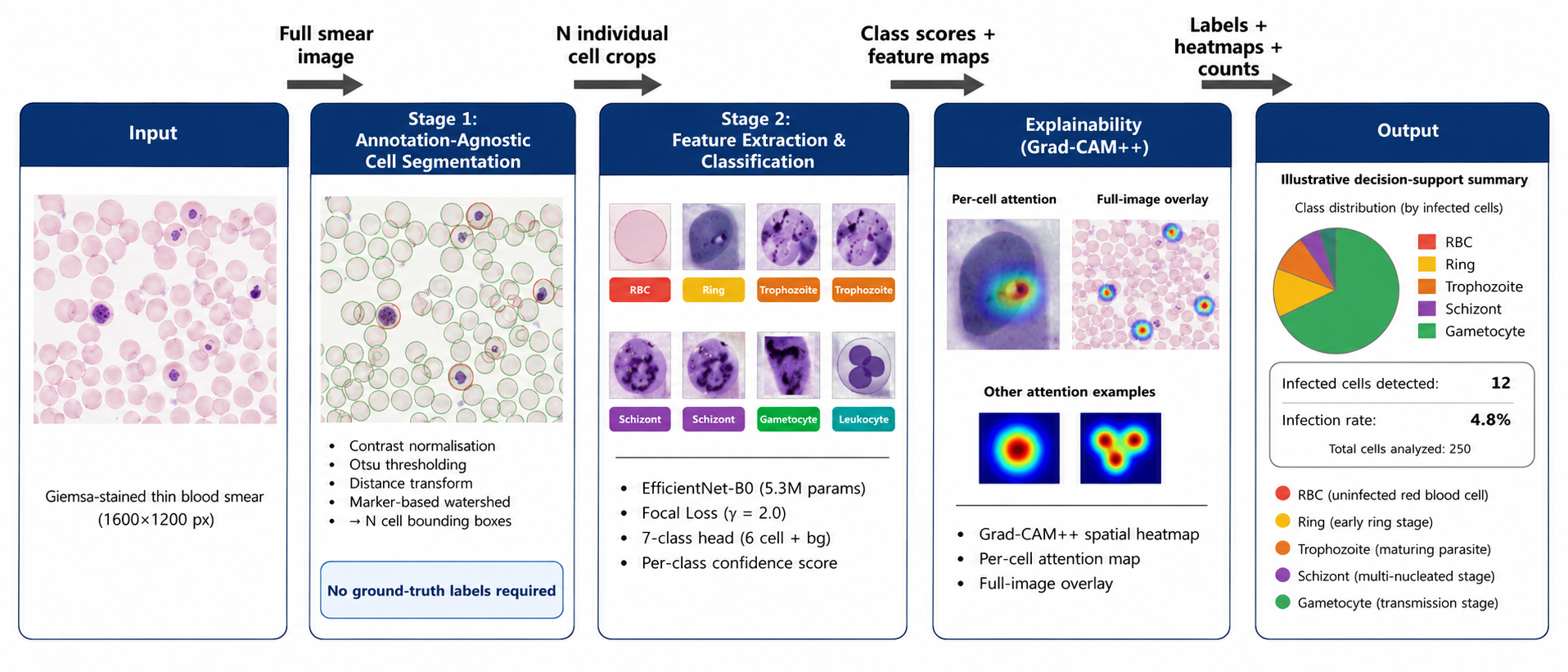}
  \caption{\malariai{} two-stage decoupled architecture. Stage~1 performs
    universal cell segmentation via distance-transform guided watershed on the raw 1600$\times$1200
    blood smear image with no annotation input. Stage~2 classifies
    each 64$\times$64 crop with EfficientNet-B0 and Focal Loss,
    followed by Grad-CAM\texttt{++} for per-cell spatial explainability.}
  \label{fig:architecture}
\end{figure}

\malariai{} follows a two-stage decoupled design, illustrated in
Figure~\ref{fig:architecture}. Stage~1 operates directly on the raw blood
smear image and produces a set of candidate cell bounding boxes using an
annotation-agnostic watershed algorithm -- no ground-truth labels are
consulted at any point. Stage~2 receives each cropped cell region and
classifies it independently using a focal-loss-trained EfficientNet-B0
classifier. A Grad-CAM\texttt{++} module then attaches a spatial heatmap to
every detected cell, enabling per-cell visual audit trails. This
decoupling deliberately separates the segmentation problem (dense,
overlapping cells in a high-resolution slide) from the classification
problem (fine-grained morphological discrimination within a single-cell
crop), allowing each stage to be optimised and evaluated independently.

\FloatBarrier

%% --------------------------
\subsubsection{Stage 1 -- Annotation-Agnostic Cell Segmentation}
\label{ssec:stage1}

Stage~1 applies a five-step computer vision pipeline to produce an ordered
set of bounding boxes $\mathcal{B} = \{(x_1^i, y_1^i, x_2^i, y_2^i)\}_{i=1}^N$
from a raw blood smear image $\mathbf{I} \in \mathbb{R}^{H \times W \times 3}$,
where $N$ is determined entirely by the image content and no ground-truth
annotation is consulted at any step.

\paragraph{Step 1: Grayscale conversion and Otsu thresholding.}
The input BGR image is converted to a single-channel grayscale image
$G \in \mathbb{R}^{H \times W}$ using the standard luminosity transform.
Otsu's method~\cite{otsu1979} selects the threshold $\tau^*$ that maximizes
inter-class variance between foreground (cells) and background:

\begin{equation}
  \tau^* = \arg\max_\tau \; \omega_0(\tau)\,\omega_1(\tau)\,
  \bigl[\mu_0(\tau) - \mu_1(\tau)\bigr]^2
  \label{eq:otsu}
\end{equation}

where $\omega_k$ and $\mu_k$ are the prior probability and mean gray level
of class $k$. The binary mask is formed as $B = \mathbf{1}[G < \tau^*]$
(inverse: cells are dark, background is bright in Giemsa-stained smears),
so foreground pixels take value 255 and background pixels take 0.

\paragraph{Step 2: Morphological opening.}
A morphological opening with a $3 \times 3$ elliptical structuring element
applied for two iterations removes thin noise filaments from the binary
mask while preserving cell bodies:

\begin{equation}
  O = (B \ominus K) \oplus K
  \label{eq:open}
\end{equation}

where $\ominus$ denotes erosion and $\oplus$ denotes dilation. The small
kernel size (3\,px) is chosen to minimize erosion of cell boundaries,
which is important for accurately locating watershed seeds at cell centers.

\paragraph{Step 3: Euclidean distance transform.}
The Euclidean distance transform $D$ assigns to each foreground pixel
$(x,y)$ its minimum distance to any background pixel:

\begin{equation}
  D(x,y) = \min_{(p,q) \notin O} \sqrt{(x-p)^2 + (y-q)^2}
  \label{eq:dist}
\end{equation}

Peaks in $D$ correspond to cell centers: a pixel at the center of a
perfectly circular cell of radius $r$ achieves $D = r$, while pixels
near the boundary achieve $D \approx 0$. In dense regions, the distance
transform naturally produces a ridge between adjacent cells, creating
the topological separator that watershed uses for instance-level splitting.

\paragraph{Step 4: Seed extraction via local maxima.}
Seeds for the watershed algorithm are extracted as local maxima of $D$
that satisfy two conditions: (i) they are separated by at least
\texttt{MIN\_DIST}~$= 12$ pixels from any higher-valued maximum, and
(ii) they exceed an absolute threshold $\tau_{\text{seed}} = 0.25 \times D_{\max}$.
The minimum distance of 12\,px (compared to the typical RBC radius of
$\sim$50\,px) ensures one seed per cell without suppressing seeds in
adjacent cells. The distance threshold of 0.25 places seeds only at
genuine cell centers, not at shallow bumps caused by staining artefacts.

\paragraph{Step 5: Watershed segmentation and bounding box extraction.}
Seeds are labeled as independent markers and watershed is applied to the
negated distance map $-D$ with the binary mask $O$ as the allowed domain:

\begin{equation}
  \mathcal{L} = \text{watershed}(-D,\; \text{markers},\; \text{mask}=O)
  \label{eq:watershed}
\end{equation}

Each labeled region $\ell$ in $\mathcal{L}$ yields a bounding box
$(x_1, y_1, x_2, y_2)$ from the extreme pixel coordinates of that region.
Regions with fewer than \texttt{MIN\_AREA}~$= 150$ pixels are discarded
as noise. A compactness parameter of 0.1 regularises elongated regions
produced by connected cell clusters, biasing watershed boundaries toward
the topological saddle points in $D$.

Figure~\ref{fig:watershed_alg} summarizes the complete Stage~1 procedure.

\begin{figure}[!t]
\centering
\setlength{\fboxsep}{3pt}%
\fbox{%
\begin{minipage}{\dimexpr\columnwidth-2\fboxsep-2\fboxrule\relax}
\footnotesize
\noindent\textbf{Algorithm 1:} Stage~1 -- Annotation-Agnostic Watershed\\
\vspace{0.3em}
\noindent\textbf{Require:} Image $\mathbf{I} \in \mathbb{R}^{H \times W \times 3}$\\[0.15em]
\textbf{Ensure:} Boxes $\mathcal{B} = \{(x_1^i, y_1^i, x_2^i, y_2^i)\}$
\vspace{0.3em}

\noindent
\begin{tabular}{@{}r@{\,}p{\dimexpr\linewidth-1.6em\relax}@{}}
1.  & $G \leftarrow \mathrm{BGR2GRAY}(\mathbf{I})$\\[2pt]
2.  & $\tau^* \leftarrow \mathrm{OtsuThreshold}(G)$\\[2pt]
3.  & $B \leftarrow \mathbf{1}[G < \tau^*]$ \quad \textit{// cells}\\[2pt]
4.  & $O \leftarrow \mathrm{MorphOpen}(B,\; k{=}3,\; \mathrm{iters}{=}2)$\\[2pt]
5.  & $D \leftarrow \mathrm{DistanceTransform}(O)$\\[2pt]
6.  & $\mathcal{S} \leftarrow \mathrm{LocalMaxima}(D,$\\
    & \quad $d_{\min}{=}12,\; \tau_{\mathrm{seed}}{=}0.25\,D_{\max})$\\[2pt]
7.  & $M \leftarrow \mathrm{LabelSeeds}(\mathcal{S})$\\[2pt]
8.  & $\mathcal{L} \leftarrow \mathrm{watershed}(-D,\; M,\; O)$;
      $\mathcal{B} \leftarrow \emptyset$\\[2pt]
9.  & \textbf{for} each $\ell \in \mathcal{L}$:\\
    & \quad\textbf{if} $|\ell| \geq 150$ \textbf{then}
      $\mathcal{B} \leftarrow \mathcal{B} \cup \{\mathrm{BBox}(\ell)\}$\\[2pt]
10.& \textbf{return} $\mathcal{B}$\\
\end{tabular}
\end{minipage}}
\caption{Stage~1 annotation-agnostic watershed segmentation (Algorithm~1).
  Parameters: \texttt{MIN\_DIST}$\!=\!12$\,px, $\tau_{\mathrm{seed}}\!=\!0.25$,
  \texttt{MIN\_AREA}$\!=\!150$\,px$^2$.}
\label{fig:watershed_alg}
\end{figure}
\FloatBarrier

\paragraph{Post-processing: oversized region filtering.}
In particularly dense regions where 3--5 cells are packed without any
visible inter-cell gap, the distance transform may not generate sufficient
seed separation to split the cluster. The resulting watershed region
spans the entire cluster and produces an oversized bounding box. Such
regions (width $> 220$\,px, height $> 220$\,px, or aspect ratio $> 2.2$)
are excluded from Stage~2 classification and flagged as ``merged cluster''
in the visual output. This prevents the EfficientNet classifier from
receiving a crop containing multiple cells - an input it was not trained
on - and avoids systematic misclassification of cluster crops as
leukocytes (the closest large-cell training class).

%% --------------------------
\subsubsection{Stage 2 -- EfficientNet-B0 Classifier with Focal Loss}
\label{ssec:stage2}

\paragraph{Crop extraction.}
For each bounding box $(x_1, y_1, x_2, y_2) \in \mathcal{B}$ produced by
Stage~1, we extract a square crop centerd on the box centroid. The crop
is resized to $64 \times 64$ pixels using bilinear interpolation. If the
centerd crop extends beyond image boundaries, the image is reflect-padded
before extraction to avoid black border artefacts. The final crop is a
uint8 RGB array normalized to ImageNet statistics ($\mu = [0.485,\,0.456,\,0.406]$,
$\sigma = [0.229,\,0.224,\,0.225]$).

\paragraph{EfficientNet-B0 architecture.}
We use EfficientNet-B0~\cite{tan2019} as the Stage~2 classification
backbone. EfficientNet applies compound scaling - simultaneously scaling
network depth, width, and input resolution by constants derived from a
neural architecture search - to achieve better parameter efficiency than
models scaled along a single axis. EfficientNet-B0 is the baseline scaling
point ($d=1.0,\;w=1.0,\;r=224$) and has 5.3M parameters, making it
deployable for CPU inference without GPU infrastructure. The final
classification head (a single Linear layer) is replaced with
$\text{Linear}(1280, 7)$ to match the seven BBBC041 categories (six cell
classes plus background). The convolutional backbone is initialized from
ImageNet~\cite{deng2009imagenet} weights to leverage low-level texture features
acquired during pretraining.

\paragraph{Focal Loss with per-class inverse-frequency weighting.}
Standard cross-entropy loss applied to the 537:1 class-imbalanced BBBC041
distribution produces a degenerate solution: the model achieves high
accuracy by predicting \textit{red blood cell} for every crop. We address
this with Focal Loss~\cite{lin2017focal} --- shown by Yeung
\etal~\cite{yeung2022} to generalise robustly across class-imbalanced medical
imaging benchmarks --- which down-weights the loss contribution of easily
classified examples:

\begin{equation}
  \mathcal{L}_{\text{focal}}(p_t) =
  -\alpha_t \,(1-p_t)^{\gamma}\,\log(p_t)
  \label{eq:focal}
\end{equation}

where $p_t$ is the model's estimated probability for the ground-truth class,
$\gamma = 2.0$ is the focusing parameter that reduces loss weight for
well-classified examples (those with $p_t \to 1$), and $\alpha_t$ is a
per-class weight. We set $\alpha_t$ by inverse-frequency normalization:

\begin{equation}
  \alpha_c = \frac{1/n_c}{\max_{c'} \; 1/n_{c'}}
  \label{eq:alpha}
\end{equation}

where $n_c$ is the number of training crops for class $c$ and the maximum
is taken over all foreground classes. This assigns $\alpha = 1.0$ to the
rarest class (leukocyte: 103 instances) and proportionally lower weights
to more frequent classes ($\alpha_{\text{RBC}} \approx 0.0013$). The
background class receives $\alpha = 0$ and contributes no gradient.
Computed weights are summarized in Table~\ref{tab:focal_weights}.

\begin{table}[!htbp]
\centering
\small
\caption{Per-class Focal Loss weights $\alpha_c$ computed from BBBC041
  training annotation counts via Equation~\ref{eq:alpha}.}
\label{tab:focal_weights}
\renewcommand{\arraystretch}{1.15}
\begin{tabular}{lrrr}
\toprule
\textbf{Class} & \textbf{Count} ($n_c$) & \textbf{$1/n_c$} & $\alpha_c$ \\
\midrule
Red blood cell  & 77,420 & $1.29 \times 10^{-5}$ & 0.0013 \\
Trophozoite     &  1,473 & $6.79 \times 10^{-4}$ & 0.0699 \\
Ring            &    353 & $2.83 \times 10^{-3}$ & 0.2918 \\
Schizont        &    179 & $5.59 \times 10^{-3}$ & 0.5754 \\
Gametocyte      &    144 & $6.94 \times 10^{-3}$ & 0.7153 \\
Leukocyte       &    103 & $9.71 \times 10^{-3}$ & 1.0000 \\
\bottomrule
\end{tabular}
\end{table}

\paragraph{Training details.}
The Stage~2 classifier is trained for 30 epochs on 79,672 pre-extracted
$64\times 64$ crops from the 966-image training split (one crop per
annotated bounding box). Pre-extraction - saving all crops as PNG files
before training begins - reduces the per-epoch I/O cost from loading
1,208 full 1600$\times$1200 images to loading 79,672 small crops,
reducing training time from $\sim$41 minutes per epoch to $\sim$50
seconds per epoch on an NVIDIA T4 GPU with batch size 128.

Training augmentation applied to each crop: random horizontal and vertical
flips (p = 0.5 each), ColorJitter with brightness $\pm 0.2$, contrast
$\pm 0.2$, and saturation $\pm 0.1$ to simulate staining variation across
laboratories. The AdamW optimizer ($lr = 3 \times 10^{-4}$,
weight decay $= 10^{-4}$) is paired with a CosineAnnealingLR schedule
(T\_max = 30, $\eta_{\min} = 10^{-6}$), smoothly decaying the learning
rate to near-zero by epoch 30 and avoiding the abrupt validation loss
increase we observe in Baseline A when step-decay is applied. The best
checkpoint is selected by maximum validation accuracy.

%% --------------------------
\subsubsection{Explainability via Grad-CAM\texttt{++}}
\label{ssec:gradcam}

\paragraph{Motivation for Grad-CAM\texttt{++} over Grad-CAM.}
Gradient-based saliency maps are among the most widely adopted tools for
building clinical trust in deep learning systems~\cite{vandervelden2022xai}.
Standard Grad-CAM~\cite{selvaraju2017} weights each spatial activation
map by the global average of its gradients with respect to the target class
score. This approach suffers from single-activation bias: when multiple
spatial locations jointly contribute to the classification (common in
small $64 \times 64$ crops where the parasite occupies most of the
receptive field), Grad-CAM over-attributes saliency to the location with
the highest single gradient value. Grad-CAM\texttt{++}~\cite{chattopadhyay2018}
addresses this with second- and third-order gradient terms that assign
more accurate per-location attribution weights.

\paragraph{Mathematical formulation.}
Let $A^k \in \mathbb{R}^{h \times w}$ denote the feature map of the
$k$-th channel at the target convolutional layer, and let $S^c$ be the
score for target class $c$ before softmax. The Grad-CAM\texttt{++} weight
for channel $k$ is:

\begin{equation}
  w_k^c = \sum_{i,j} \alpha_{kij}^c \cdot \text{ReLU}
  \!\left(\frac{\partial S^c}{\partial A_{ij}^k}\right)
  \label{eq:gcampp_weight}
\end{equation}

where the pixel-wise coefficient $\alpha_{kij}^c$ is:

\begin{equation}
  \alpha_{kij}^c = \frac{
    \left(\dfrac{\partial^2 S^c}{\partial (A_{ij}^k)^2}\right)
  }{
    2\,\dfrac{\partial^2 S^c}{\partial (A_{ij}^k)^2}
    + \displaystyle\sum_{a,b} A_{ab}^k \cdot
    \dfrac{\partial^3 S^c}{\partial (A_{ij}^k)^3} + \varepsilon
  }
  \label{eq:alpha_gcam}
\end{equation}

In practice, denoting first-, second-, and third-order gradient tensors
as $G^{(1)} = \partial S^c / \partial A^k$, $G^{(2)} = (G^{(1)})^2$,
and $G^{(3)} = (G^{(1)})^3$, Equation~\ref{eq:alpha_gcam} becomes:

\begin{equation}
  \alpha^c = \frac{G^{(2)}}{2\,G^{(2)}
  + \bigl(A \cdot G^{(3)}\bigr)_{\text{sum}} + \varepsilon}
  \label{eq:alpha_gcam_compact}
\end{equation}

The final heatmap is:

\begin{equation}
  L^c_{\text{Grad-CAM++}} = \text{ReLU}
  \!\left(\sum_k w_k^c \cdot A^k\right)
  \label{eq:gcampp_cam}
\end{equation}

upsampled bilinearly to the input resolution ($64 \times 64$) and
normalized to $[0, 1]$.

\paragraph{Hook target.}
Hooks are registered on \texttt{model.features[-1]}, the final MBConv
block of EfficientNet-B0, producing feature maps of shape $[B, 1280, 2, 2]$
for $64\times 64$ inputs. This layer has a receptive field spanning most
of the crop, making it the most informative layer for attribution.

\paragraph{Output views.}
Two visualizations are produced per blood smear:
(i) a \textit{Grad-CAM\texttt{++} crop gallery} showing each detected
parasite as a 160$\times$160 pixel overlay of the original crop and
heatmap (jet colormap, $\alpha = 0.5$), labeled with class name and
confidence; and
(ii) a \textit{full-image Grad-CAM\texttt{++} overlay} that splats each
parasite's heatmap back onto the full 1600$\times$1200 smear at the
bounding box coordinates, using a spatial mask so only parasite-detected
regions are heatmap-colored and the remainder of the image is unchanged.

\FloatBarrier
%% --------------------------
\subsection{Baseline Method (Faster R-CNN)}
\label{ssec:baseline}

Baseline A is a Faster R-CNN~\cite{ren2015} with a ResNet-50 Feature
Pyramid Network (FPN) backbone pretrained on MS-COCO. The standard 91-class
classification head is replaced with a 7-class \texttt{FastRCNNPredictor}.
The FPN generates multi-scale feature maps at strides \{4, 8, 16, 32, 64\},
enabling detection of both large cells (\eg, leukocytes, median area
$\approx$14,000\,px$^2$) and small ring-stage nuclei ($\approx$3,500\,px$^2$).

\paragraph{Training protocol.}
The model is trained for 80 epochs on the 966-image training split using
SGD ($lr = 5 \times 10^{-3}$, momentum $= 0.9$, weight decay $= 10^{-4}$)
with a MultiStepLR schedule ($\gamma = 0.1$ at epoch 50) and gradient
clipping (max norm 10.0). Batch size 4 on an NVIDIA T4 GPU. Best checkpoint
selected by minimum validation loss.

\begin{figure}[!htbp]
  \centering
  \includegraphics[width=\linewidth]{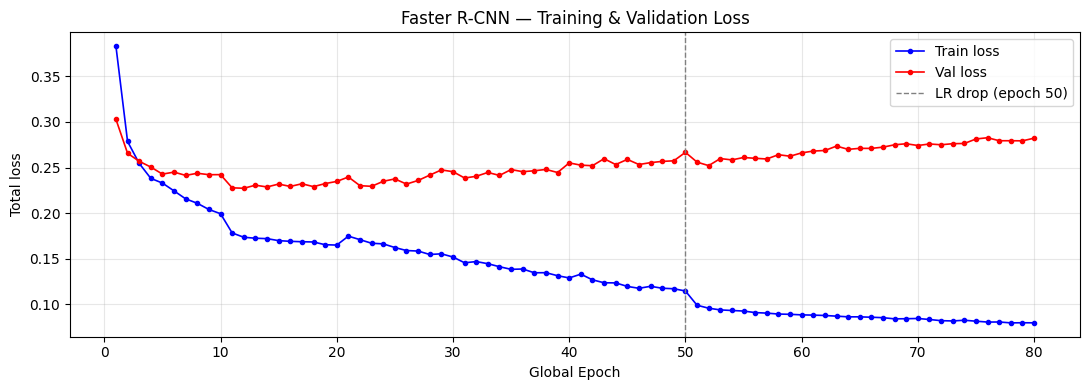}
  \caption{Faster R-CNN (Baseline A) training and validation loss over
    80 epochs. Validation loss plateaus at $\approx 0.23$ from epoch~8
    and increases after the learning rate drop at epoch~50 - the
    empirical signature of annotation incompleteness (Problem~P1).
    Best checkpoint: epoch~23 (val.\ loss 0.2294).}
  \label{fig:frcnn_loss}
\end{figure}

\paragraph{Structural training ceiling.}
Figure~\ref{fig:frcnn_loss} shows the training and validation loss curves.
Validation loss plateaus at $\approx 0.23$ from epoch~8, while training
loss continues to fall. After the learning rate drop at epoch~50, training
loss decreases to 0.08 while validation loss \emph{increases} to 0.28.
This divergence begins before any opportunity for conventional overfitting
and is the empirical signature of Problem~P1: the model correctly detects
unannotated cells but is penalised by the closed-world loss function. The
best checkpoint is epoch~23 (val.\ loss 0.2294); 57 subsequent epochs
produce no improvement.

\begin{figure}[!htbp]
  \centering
  \includegraphics[width=\linewidth]{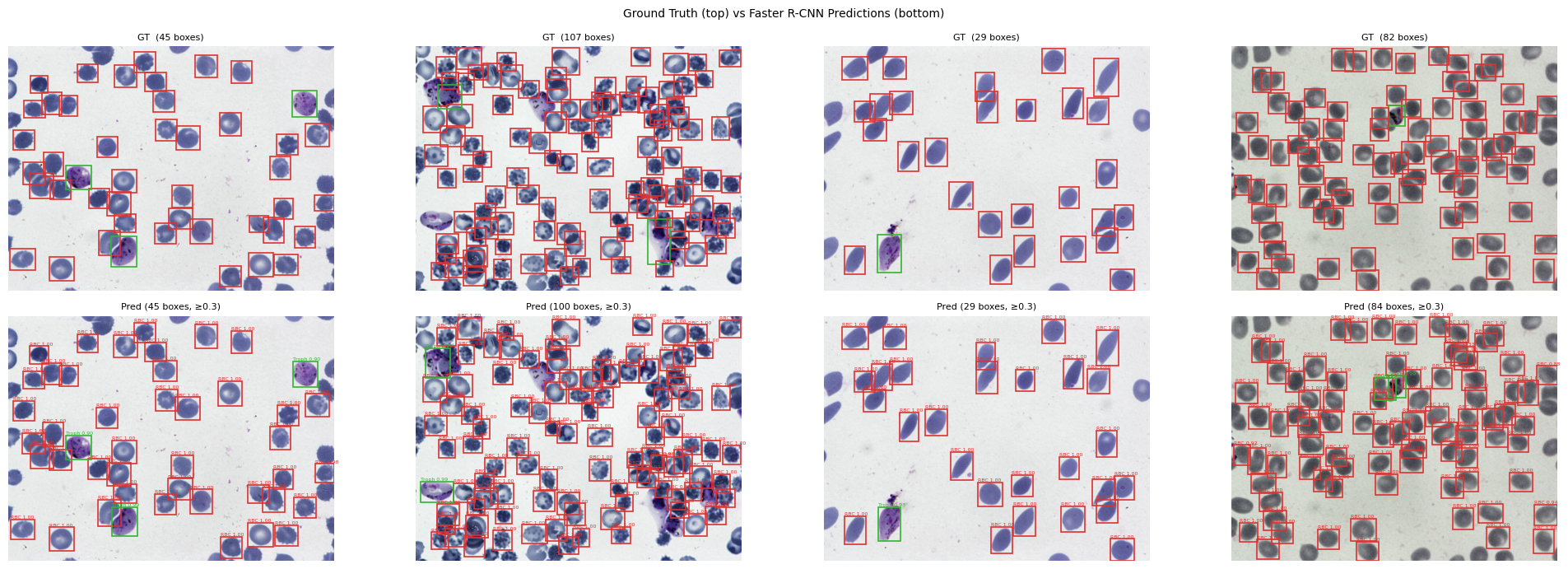}
  \caption{Ground truth (top) versus Faster R-CNN Baseline A predictions
    (bottom) on four validation images. Color coding: red = red blood cell;
    remaining colors = parasitic stages and leukocyte. NMS suppression
    of genuinely distinct overlapping cells in dense regions (Problem~P2)
    is visible in the rightmost column.}
  \label{fig:frcnn_predictions}
\end{figure}

Figure~\ref{fig:frcnn_predictions} shows qualitative predictions on four
validation images, comparing ground truth against Baseline A's output
directly. NMS suppression of genuinely distinct, overlapping cells in
dense regions (Problem~P2) is visible in the rightmost column, where
adjacent cells that should be detected separately collapse into a
single retained box.

\FloatBarrier
%% --------------------------
\subsection{Baseline B (YOLOv8)}
\label{ssec:baseline_b}

While Baseline A establishes a 2015-era detection reference, several
recent malaria detection systems benchmark against modern single-stage
detectors~\cite{yolov4malaria2024,issah2026}. To situate \malariai{}
against current state-of-the-art detection architecture, not only an
older two-stage detector, we additionally train YOLOv8s~\cite{yolov8}
as Baseline B on the identical 966/242 image-level split used for
Baseline A (Section~\ref{ssec:baseline}), so both baselines are scored
on exactly the same held-out validation images.

\paragraph{Training protocol.}
YOLOv8s is trained from COCO-pretrained weights for 80 epochs at
$1024\times1024$ input resolution, batch size 16, on an NVIDIA T4 GPU
(Kaggle Notebooks), using the same 6-class label set as Baseline A.
Predictions below confidence 0.3 are discarded, matching Baseline A's
evaluation protocol (Section~\ref{ssec:metrics}).

\begin{figure}[!htbp]
  \centering
  \includegraphics[width=\linewidth]{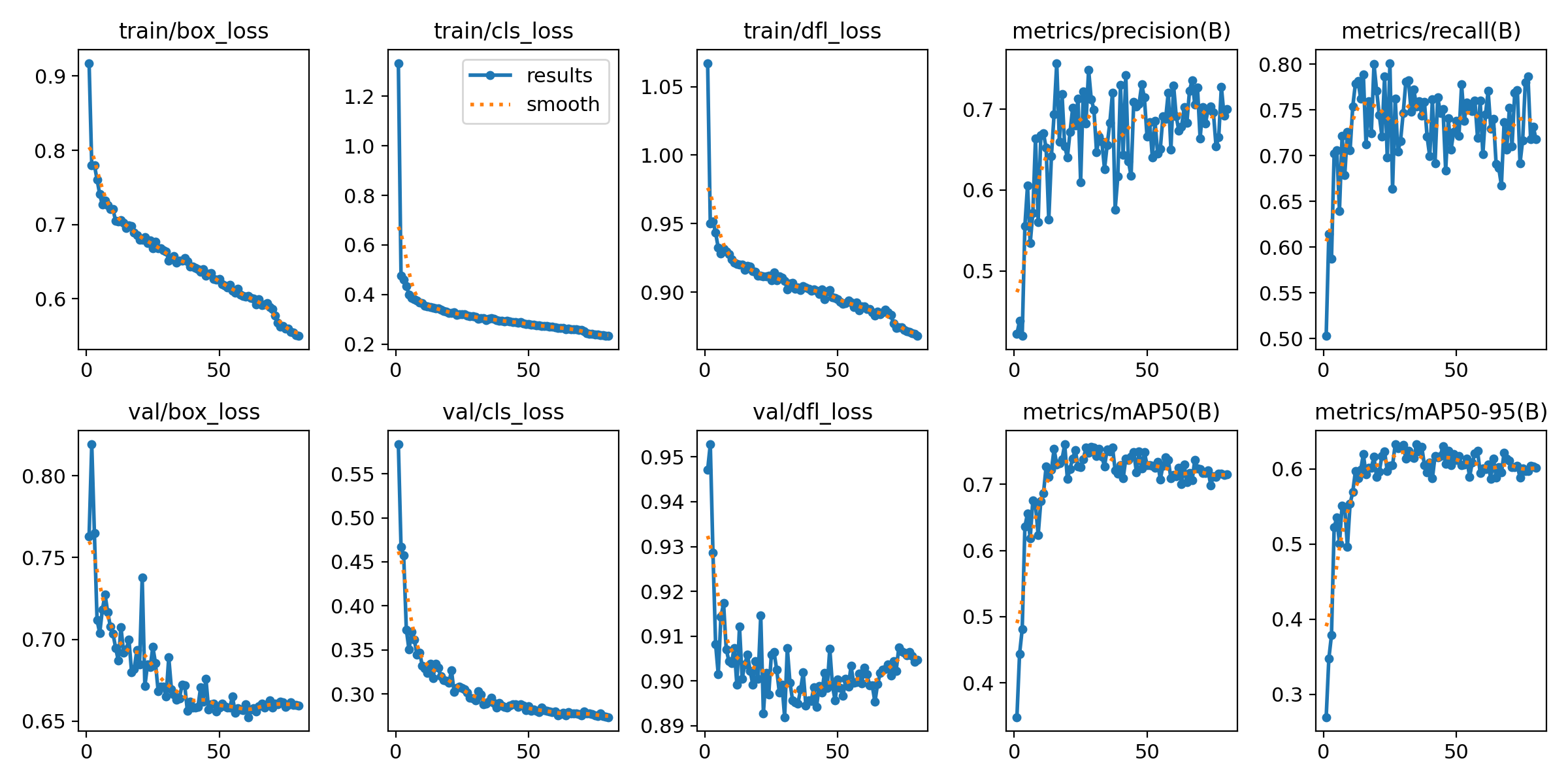}
  \caption{YOLOv8s (Baseline B) training and validation loss and mAP
    curves over 80 epochs on the same 966/242 split as Baseline A.}
  \label{fig:yolo_loss}
\end{figure}

\begin{figure}[!htbp]
  \centering
  \includegraphics[width=\linewidth]{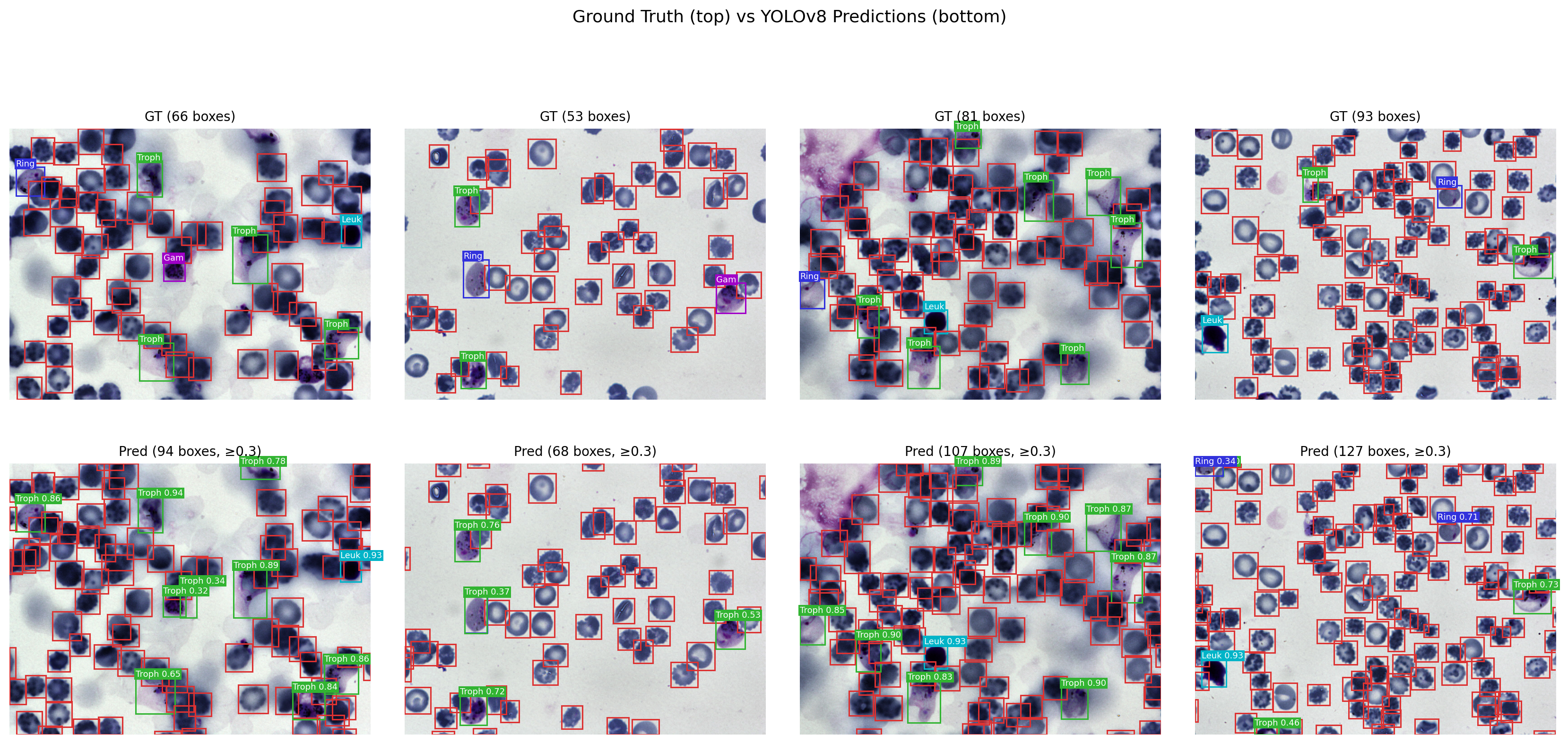}
  \caption{Ground truth (top) versus YOLOv8s Baseline B predictions
    (bottom) on four validation images, selected for class diversity.
    Only parasitic-stage and leukocyte boxes are labeled; red blood
    cell boxes are shown unlabeled for readability given their high
    density per image.}
  \label{fig:yolo_predictions}
\end{figure}

Figure~\ref{fig:yolo_predictions} shows the corresponding qualitative
predictions, selected for class diversity so both common (red blood
cell) and rare parasitic-stage detections are visible side by side with
ground truth.

\begin{figure}[!htbp]
  \centering
  \includegraphics[width=\linewidth]{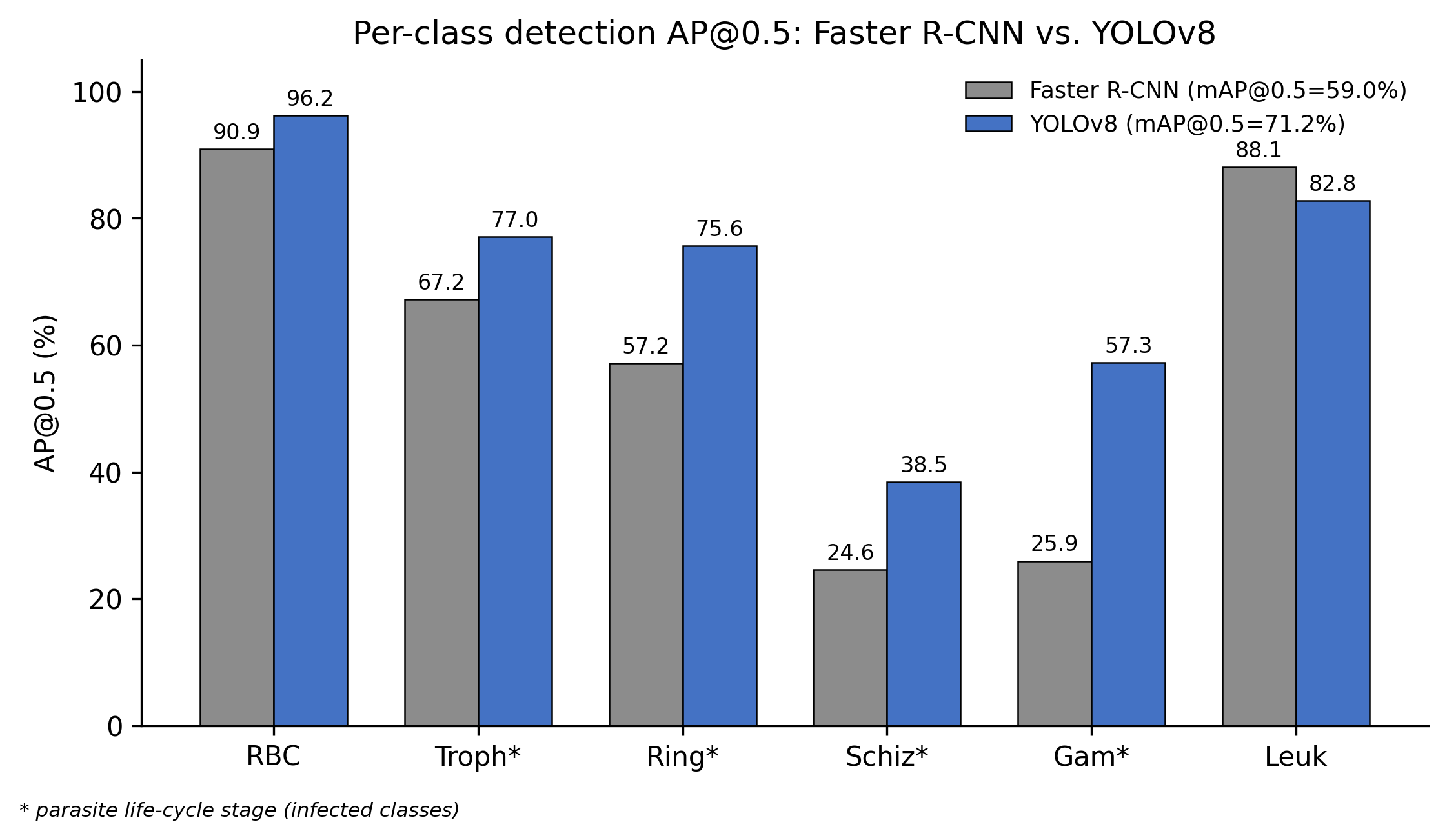}
  \caption{Per-class detection AP@0.5: Faster R-CNN (Baseline A) versus
    YOLOv8s (Baseline B) on the identical 242-image validation split.
    YOLOv8s improves mAP@0.5 from 58.99\% to 71.24\% overall, with the
    largest gains on the rare parasitic stages (schizont: 24.57\%
    $\rightarrow$ 38.45\%; gametocyte: 25.95\% $\rightarrow$ 57.27\%).}
  \label{fig:yolo_vs_frcnn}
\end{figure}

Baseline B substantially outperforms Baseline A as a detector
(Figure~\ref{fig:yolo_vs_frcnn}), confirming that a modern architecture
narrows - but does not close - the rare-class detection gap that
motivates \malariai{}'s decoupled design. Section~\ref{ssec:stage2_results}
places Baseline B alongside Baseline A and Pipeline B in a full
per-class comparison.

\FloatBarrier
%% --------------------------
\subsection{Evaluation Metrics}
\label{ssec:metrics}

\paragraph{Cell recovery rate (Pipeline B Stage 1).}
Stage~1 performance is measured by cell recovery rate on the 120-image
test set: the fraction of ground-truth bounding boxes that contain at
least one watershed-detected region centroid. This annotation-agnostic
metric directly quantifies P1 resilience. Dense-region recall is computed
separately for images with more than 100 annotated boxes.

\paragraph{Classification metric (Pipeline B Stage 2).}
Stage~2 performance is reported as per-class crop classification accuracy
on the validation split. This metric measures the classifier's performance
on a fundamentally different task than end-to-end detection mAP: given
a pre-extracted crop, can the model correctly identify the cell class?
\textbf{These two metrics measure different tasks and are not directly
comparable.} Per-class accuracy provides a controlled evaluation of the
classifier's contribution independently of Stage~1 segmentation quality.

\paragraph{Detection metric (Baseline A).}
Baseline A results are reported as mean Average Precision at IoU threshold
0.5 (mAP@0.5) using \texttt{torchmetrics.MeanAveragePrecision} with
\texttt{class\_metrics=True}. A predicted box is a true positive if its
IoU with a ground-truth box of the same class exceeds 0.5. Predictions
with confidence below 0.3 are discarded. Per-class AP is reported for
all six BBBC041 categories.

%% ============================================================================
%% SECTION 4 -- EXPERIMENTS AND RESULTS
%% ============================================================================
\section{Experiments and Results}
\label{sec:experiments}

\subsection{Implementation Details}
\label{ssec:impl}

\paragraph{Hardware and software.}
All training is conducted on an NVIDIA T4 GPU (16\,GB VRAM) via Kaggle
Notebooks. Inference for the deployed application runs on CPU only.
The codebase is implemented in PyTorch 2.x with \texttt{torchvision} for
Baseline A and \texttt{scikit-image}~\cite{scikitimage2014} for Stage~1 watershed operations.

\paragraph{Baseline A (Faster R-CNN).}
Architecture: ResNet-50~\cite{he2016resnet} FPN~\cite{lin2017fpn}, MS-COCO~\cite{lin2014coco} pretrained, 7-class head. Optimizer:
SGD ($lr = 5 \times 10^{-3}$, momentum 0.9, weight decay $10^{-4}$).
Schedule: MultiStepLR at epoch 50 ($\gamma = 0.1$). Batch size 4. Epochs:
80. Gradient clipping: max norm 10.0. Best checkpoint: epoch 23
(val.\ loss 0.2294). Training time: $\approx$5.5 hours.

\paragraph{Baseline B (YOLOv8s).}
Architecture: YOLOv8s~\cite{yolov8}, COCO-pretrained, 6-class head
(Section~\ref{ssec:baseline_b}). Optimizer: SGD with automatic
learning-rate selection (Ultralytics default). Input resolution
$1024\times1024$. Batch size 16. Epochs: 80. Trained and evaluated on
the identical 966/242 image-level split (seed 42) as Baseline A.
Training time: $\approx$1 hour.

\paragraph{Pipeline B (\malariai{}).}
\textit{Stage 1 parameters:} \texttt{MIN\_AREA}~$= 150$\,px$^2$,
\texttt{MIN\_DIST}~$= 12$\,px, \texttt{OPEN\_KSIZE}~$= 3$\,px,
\texttt{DIST\_THRESH}~$= 0.25$, confidence threshold $= 0.40$ on
parasite predictions.
\textit{Stage 2 parameters:} EfficientNet-B0, ImageNet pretrained; input
$64\times 64$; Focal Loss ($\gamma = 2.0$, per-class $\alpha$ from
Table~\ref{tab:focal_weights}); AdamW ($lr = 3 \times 10^{-4}$,
weight decay $10^{-4}$); CosineAnnealingLR (T\_max = 30,
$\eta_{\min} = 10^{-6}$); batch size 128; 30 epochs. Training time:
$\approx$50\,s/epoch ($\approx$25 minutes total) after crop pre-extraction.
\textit{Grad-CAM\texttt{++} target layer:} the standard choice for
EfficientNet-style backbones -- the final convolutional block
(\texttt{features[-1]}) -- collapses a $64\times 64$ input to
approximately a $2\times 2$ feature map under EfficientNet-B0's stride
pattern, too coarse to localize within a tightly-margined crop. We
instead hook \texttt{features[5]} ($4\times 4$ spatial resolution),
selected via a layer ablation (Section~\ref{ssec:gradcam_quant})
comparing candidate layers by how far their energy-in-box ratio exceeds
a geometric chance baseline.

\paragraph{Learned segmentation baseline (U-Net).}
Architecture: standard 4-level encoder-decoder U-Net~\cite{ronneberger2015unet}
(base width 32, BCE + Dice loss). Since BBBC041 provides bounding-box
annotations only (no pixel-level masks), training targets are
synthetic ellipse masks filled inside each ground-truth box. Images are
downsampled to $512\times 384$ for training and inference. Optimizer:
Adam, batch size 8, 25 epochs, trained on the same 967-image training
split used elsewhere (never exposed to the 120-image test set).
Predicted probability maps are converted to boxes either by (i) a
global threshold followed by connected-component labelling, matching
the same oversized-region filter used for Stage~1
(Section~\ref{ssec:stage1}), or (ii) the same distance-transform
watershed splitting Stage~1 already applies to raw pixel intensity,
applied instead to the U-Net probability map (Section~\ref{ssec:unet_baseline}).

\FloatBarrier
\subsection{Stage 1 Performance}
\label{ssec:stage1_perf}

\subsubsection{Evaluation: Cell Recovery Rate}
\label{ssec:stage1_results}

Table~\ref{tab:stage1} reports Stage~1 (Annotation-Agnostic Cell Segmentation, Section~\ref{ssec:stage1}) performance on the held-out
120-image test set (5,917 valid ground-truth boxes). A watershed region
is counted as recovering a ground-truth box if any Stage~1 centroid falls
within that box.

\begin{table}[!htbp]
\centering
\small
\caption{Stage~1 (watershed) cell recovery on the NIH BBBC041 120-image
  test set. A ground-truth box is ``recovered'' if any watershed centroid
  falls within its boundary. Dense images have $\geq 100$ annotated boxes.}
\label{tab:stage1}
\renewcommand{\arraystretch}{1.15}
\begin{tabular}{lc}
\toprule
\textbf{Metric} & \textbf{Value} \\
\midrule
Test images evaluated              & 120 \\
Ground-truth boxes (valid)         & 5,917 \\
GT boxes recovered                 & 4,494 \\
\textbf{Cell recovery rate}        & \textbf{75.95\%} \\
\midrule
Dense images ($\geq 100$ GT boxes) & 2 \\
Dense-region recall                & 50.94\% \\
\bottomrule
\end{tabular}
\end{table}

The overall cell recovery rate of 75.95\% is achieved without any
annotation input, directly validating Contribution~C1. The dense-region
recall of 50.94\% is computed over only two qualifying images ($\geq 100$
GT boxes in the 120-image test set), limiting its statistical significance;
we treat this as a preliminary indicator and discuss it further in
Section~\ref{sec:discussion}.

%% -------------------------
\subsubsection{Localization: Alternative Metrics}
\label{ssec:stage1_alt_metrics}

The standard IoU@0.5 criterion was designed for rectangular proposal-based
detectors and systematically penalises watershed segmentation: because the
watershed algorithm grows organic, blob-shaped regions rather than fitting
tight rectangular boxes, a region may correctly identify the spatial location
of a cell yet still fail the strict IoU@0.5 threshold.  To quantify how much
of the apparent recall gap is attributable to boundary mis-alignment rather
than genuine missed detections, we report four complementary localisation
metrics computed by the updated evaluation framework.\footnote{\texttt{src/pipeline\_b\_v2/e2e\_eval.py}, available in the code repository (see Code and Demo Availability).}

\begin{enumerate}[leftmargin=*, label=\textbf{M\arabic*.}]

  \item \textbf{Centroid-in-box recall.}  A ground-truth box is considered
  recovered if the centroid of any watershed region falls within its boundary,
  regardless of overlap area.  This criterion tests spatial localisation
  without penalising non-rectangular region boundaries.

  \item \textbf{Relaxed IoU recalls.}  Recall is computed at IoU thresholds
  of 0.25 and 0.30 in addition to the standard 0.50. These thresholds are
  selected on geometric grounds: for a circular cell approximated by a square
  bounding box, a watershed blob that correctly encloses the cell body but
  extends slightly outside the annotation boundary can easily achieve
  IoU~0.30 while failing IoU~0.50. The relaxed thresholds therefore isolate
  boundary mis-alignment from genuine missed detections.

  \item \textbf{Infected-cell sensitivity.}  Recall is computed exclusively
  over the parasitised ground-truth boxes (ring, trophozoite, schizont,
  gametocyte on BBBC041; all four \textit{Plasmodium} species on MP-IDB).
  In clinical malaria diagnosis, sensitivity is prioritized over specificity:
  a missed infected cell (false negative) carries critical clinical consequences,
  including delayed treatment and increased patient risk, whereas false positives
  can be resolved by human review~\cite{who2023}.

  \item \textbf{Biological localisation recall.}  A watershed region is
  counted as a hit if its centroid lies within $0.5 \times d_{\text{GT}}$
  pixels of the ground-truth centroid, where $d_{\text{GT}}$ is the
  Euclidean diagonal of the GT box.  For a typical BBBC041 RBC
  ($65 \times 65$\,px, $d_{\text{GT}} \approx 92$\,px) this allows
  $\approx 46$\,px displacement - roughly one cell radius.  This criterion
  is sufficient for a pathologist to verify a detection location visually.

\end{enumerate}

Throughout this paper, two Stage~1 configurations are evaluated:
\textbf{Stage~1 v1} uses the original Otsu global thresholding with a fixed
distance-ratio seed threshold, and \textbf{Stage~1 v2} replaces this with
CLAHE contrast normalization and resolution-aware \texttt{peak\_local\_max}
seeding (introduced and compared in Section~\ref{ssec:stage1_v2_ablation}).
Table~\ref{tab:alt_metrics} reports these metrics on the 120-image BBBC041
test set (Stage~1 v1, standard watershed).  The values should be read
alongside the IoU@0.5 recall of 66.88\% from Table~\ref{tab:e2e_bbbc041}.

\begin{table}[!htbp]
\centering
\small
\caption{Alternative localisation metrics for Stage~1 on the BBBC041
  120-image test set (Stage~1 v1, 5,917 GT boxes, 7,704 watershed regions).
  All metrics evaluate spatial detection only; Stage~2 classification is not
  involved.  Values in parentheses are percentages.}
\label{tab:alt_metrics}
\renewcommand{\arraystretch}{1.2}
\begin{tabular}{lcc}
\toprule
\textbf{Metric} & \textbf{TP / GT} & \textbf{Recall} \\
\midrule
\multicolumn{3}{l}{\textit{Standard: IoU $\geq$ 0.50}} \\
\quad Recall @ IoU $\geq 0.50$ & 3,957 / 5,917 & 66.88\% \\
\midrule
\multicolumn{3}{l}{\textit{M1: Centroid-in-GT-box (boundary-independent spatial localization)}} \\
\quad Centroid-in-box recall    & 4,677 / 5,917 & \textbf{79.04\%} \\
\midrule
\multicolumn{3}{l}{\textit{M2: Relaxed IoU thresholds}} \\
\quad Recall @ IoU $\geq 0.25$ & -- / 5,917 & \textbf{78.08\%} \\
\quad Recall @ IoU $\geq 0.30$ & -- / 5,917 & 76.80\% \\
\quad Recall @ IoU $\geq 0.50$ & 3,957 / 5,917 & 66.88\% \\
\midrule
\multicolumn{3}{l}{\textit{M3: Infected-cell sensitivity ($n = 303$ parasitised GT cells)}} \\
\quad Sensitivity @ IoU $\geq 0.25$ & -- / 303 & \textbf{82.84\%} \\
\quad Sensitivity @ IoU $\geq 0.30$ & -- / 303 & 80.20\% \\
\quad Sensitivity @ IoU $\geq 0.50$ & -- / 303 & 66.34\% \\
\midrule
\multicolumn{3}{l}{\textit{M4: Biological localisation (WS centroid $\leq 0.5 \times d_{\text{GT}}$ from GT centroid)}} \\
\quad Bio-localisation recall   & 4,704 / 5,917 & \textbf{79.50\%} \\
\bottomrule
\end{tabular}
\end{table}

Table~\ref{tab:alt_metrics} reveals a consistent pattern across all four
metrics: the IoU@0.5 recall of 66.88\% understates the true spatial
detection capability of Stage~1.

\textbf{Boundary mis-alignment penalty.}  The centroid-in-box recall
of 79.04\% (M1) and the bio-localisation recall of 79.50\% (M4) are
both approximately 12~percentage points above the IoU@0.5 figure.  This
gap measures the cost of insisting on rectangular boundary overlap: a
watershed region whose organic boundary partially extends beyond the
annotation rectangle loses the IoU@0.5 test even though its center is
correctly placed inside the cell.

\textbf{Relaxed boundary criterion.}  At IoU~$\geq 0.25$, recall
rises to 78.08\% (M2), closely matching the boundary-free metrics.  The
large recall jump from IoU@0.50 (66.88\%) to IoU@0.25 (78.08\%) confirms
that most ``missed'' detections at the strict threshold are in fact
correctly localized cells whose region-to-box overlap falls in the
0.25--0.50 range - a direct consequence of irregular watershed
boundaries meeting rectangular annotations.

\textbf{Infected-cell sensitivity.}  The clinically decisive finding is in
M3: at IoU~$\geq 0.25$, Stage~1 recovers 82.84\% of the 303~parasitised
GT boxes, compared to 66.34\% at IoU~$\geq 0.50$.  The 16.5-point
gap for infected cells specifically shows that the majority of missed
parasites at strict IoU are spatially found by the watershed - the
detector's localisation is sufficient for clinical purposes, even when its
region boundary does not perfectly match the rectangular annotation.

Taken together, these metrics support the interpretation that Stage~1 is
a strong spatial localizer: approximately 79--83\% of all cells (and of
infected cells specifically) are correctly identified spatially, with the
remaining gap attributable to annotation boundary conventions rather than
genuine missed detections.

\FloatBarrier
\subsubsection{Learned Segmentation Baseline (U-Net)}
\label{ssec:unet_baseline}

Stage~1's watershed algorithm requires no training and no annotation
input, but this leaves open a natural question: how does it compare to
a conventionally-trained supervised segmentation network on the same
task? We trained a standard U-Net (Section~\ref{ssec:impl}) on synthetic
ellipse masks derived from the same ground-truth boxes and evaluated it
on the identical 120-image test set used for Stage~1
(Table~\ref{tab:stage1}), so the comparison in
Table~\ref{tab:unet_vs_watershed} uses the same images, the same
5,917 ground-truth boxes, and the same IoU~$\geq 0.5$ matching
convention as Table~\ref{tab:stage1}.

\begin{table}[!htbp]
\centering
\small
\caption{U-Net segmentation baseline vs.\ Stage~1 watershed, evaluated
  on the identical 120-image BBBC041 test set (5,917 GT boxes,
  IoU~$\geq 0.5$ matching). ``Naive threshold'' converts U-Net's
  predicted probability map to boxes via global thresholding and
  connected-component labelling; ``+ watershed-split'' instead applies
  the same distance-transform watershed splitting Stage~1 uses, but to
  the U-Net probability map rather than raw pixel intensity.}
\label{tab:unet_vs_watershed}
\renewcommand{\arraystretch}{1.2}
\begin{tabular}{lccc}
\toprule
\textbf{Method} & \textbf{Recall} & \textbf{Precision} & \textbf{F1} \\
\midrule
Stage~1 watershed (raw pixel intensity) & \textbf{66.88\%} & 51.36\% & 0.581 \\
U-Net, naive threshold                   & 25.28\%          & 79.41\% & 0.384 \\
U-Net + watershed-split                  & 62.89\%          & \textbf{83.28\%} & \textbf{0.717} \\
\bottomrule
\end{tabular}
\end{table}

\begin{figure}[!htbp]
  \centering
  \includegraphics[width=0.9\columnwidth]{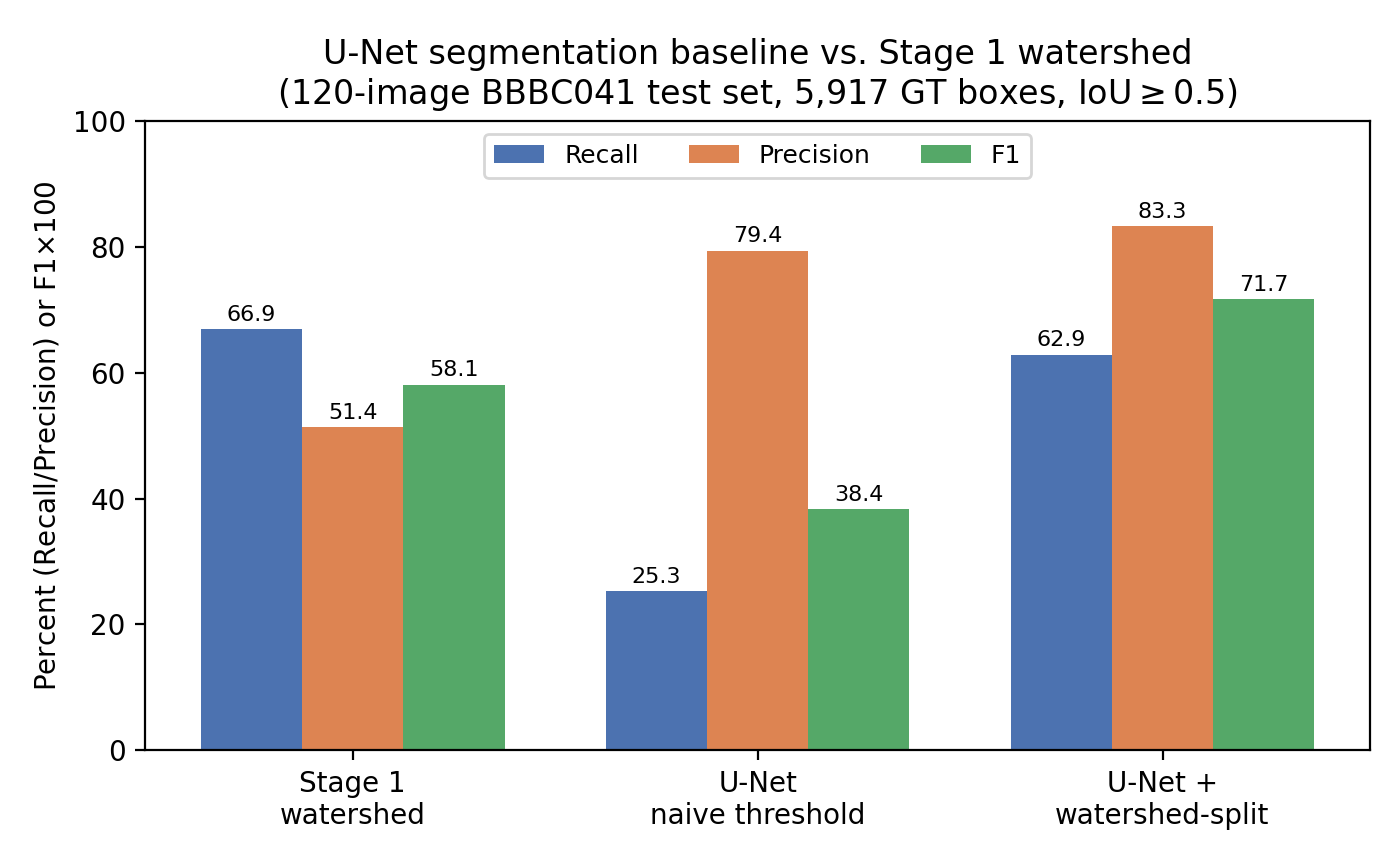}
  \caption{U-Net segmentation baseline vs.\ Stage~1 watershed, same
    120-image BBBC041 test set and IoU~$\geq 0.5$ matching as
    Table~\ref{tab:stage1}. Naive thresholding of the U-Net probability
    map underperforms watershed on recall; applying the same
    watershed-splitting technique to the U-Net probability map instead
    of raw pixel intensity closes most of the gap and wins on precision
    and F1.}
  \label{fig:unet_vs_watershed}
\end{figure}

\textbf{Naive thresholding under-segments.} With simple global
thresholding, U-Net recovers only 25.28\% of ground-truth cells --
substantially below watershed's 66.88\% -- despite comparatively high
precision (79.41\%). This is not primarily a detection failure: sweeping
the threshold from 0.2 to 0.9 shows recall rising monotonically while
precision stays essentially flat and the number of predicted regions
grows -- the specific signature of adjacent cells being merged into
single blobs at low thresholds and separating only as the threshold
tightens, rather than the model progressively detecting more real
cells. Diagnostic inspection confirms this directly: at
threshold~$=0.5$, U-Net produces an average of 28.9 connected
components per image against an average of 67.3 ground-truth boxes per
image -- roughly one predicted region for every 2.3 true cells,
consistent with systematic merging of touching cells rather than missed
detections.

\textbf{Watershed-style splitting recovers most of the gap, and wins on
precision.} Rather than treat this as a ceiling on U-Net's usefulness,
we applied Stage~1's own remedy for exactly this failure mode --
distance-transform-guided watershed splitting -- directly to U-Net's
predicted probability map in place of a single global threshold.
This closes most of the recall gap (62.89\% vs.\ 66.88\%, within 4
points of Stage~1) while substantially exceeding Stage~1 on both
precision (83.28\% vs.\ 51.36\%) and F1 (0.717 vs.\ 0.581). The
precision gain is notable given both methods rely on the identical
splitting algorithm at this stage: it indicates that the U-Net's
learned probability map is a cleaner signal to seed watershed from than
raw pixel intensity, producing fewer spurious low-confidence regions
even though its raw recall remains marginally lower.

\textbf{Interpretation.} We read this as an honest, non-oversold
finding rather than a claim that supervised segmentation categorically
outperforms Stage~1's annotation-free approach. Naive thresholding of a
learned segmentation mask is not sufficient on its own -- the same
touching-cell problem that motivates watershed's use in Stage~1 also
afflicts a trained U-Net, and connected-component labelling alone does
not solve it. What the experiment does show is that the choice of
\textit{post-processing step} (splitting merged regions), not merely
the choice of \textit{signal source} (raw intensity vs.\ a learned
mask), is the dominant factor governing detection quality in this
task, and that Stage~1's own watershed-splitting design generalizes
well when applied to a different upstream signal.

\FloatBarrier
\subsection{Stage 2 Performance -- Crop Classification Results}
\label{ssec:stage2_results}

\begin{figure}[!htbp]
  \centering
  \includegraphics[width=\linewidth]{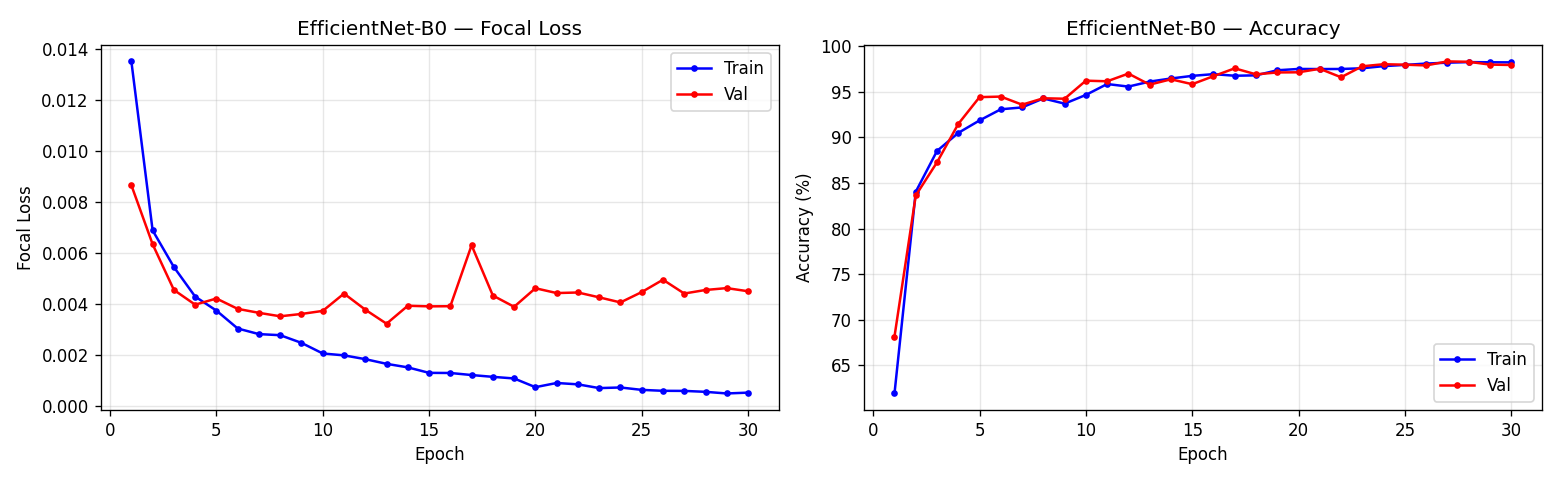}
  \caption{Pipeline B Stage~2 (EfficientNet-B0) training and validation
    loss and accuracy curves over 30 epochs. Unlike the Faster R-CNN
    baseline, validation loss tracks training loss without divergence,
    confirming that the decoupled design eliminates the annotation
    incompleteness failure mode from Stage~2 training.
    Best validation accuracy: 98.36\% at epoch~27.}
  \label{fig:stage2_loss}
\end{figure}

Figure~\ref{fig:stage2_loss} shows the Stage~2 training and validation
loss and accuracy curves over 30 epochs. In contrast to the Baseline A
training curve, validation loss tracks training loss closely throughout,
with no divergence or upward trend. This is the expected behavior under
our decoupled design: Stage~2 is trained on crops extracted from
ground-truth bounding boxes, which are complete and unambiguous labels.
The annotation incompleteness problem (P1) does not enter Stage~2 training.

Table~\ref{tab:stage2_acc} reports per-class classification accuracy on
the validation split. The overall accuracy of 98.36\% is largely driven
by the dominant red blood cell class (98.82\%). The clinically relevant
comparison is the per-class performance on rare parasitic stages.

\begin{table}[!htbp]
\centering
\small
\caption{Stage~2 (EfficientNet-B0) per-class crop classification accuracy
  on the validation split, alongside both detection baselines evaluated
  on the identical 242-image validation split. \textit{Note:} Crop
  accuracy and detection AP@0.5 are different metrics measuring
  different tasks and are not directly comparable (Section~\ref{ssec:metrics}).
  $\dagger$ = rare class ($<200$ training instances).}
\label{tab:stage2_acc}
\renewcommand{\arraystretch}{1.2}
\begin{tabular*}{\textwidth}{@{\extracolsep{\fill}}lccc@{}}
\toprule
\textbf{Class} & \textbf{Pipeline B} & \textbf{Baseline A} & \textbf{Baseline B} \\
               & \textit{Crop Accuracy} & \textit{AP@0.5} & \textit{AP@0.5 (YOLOv8s)} \\
\midrule
Red blood cell        & 98.82\%  & 90.94\% & 96.24\% \\
Leukocyte             & 100.00\% & 88.12\% & 82.79\% \\
Trophozoite           & 83.71\%  & 67.22\% & 77.05\% \\
Ring                  & 77.78\%  & 57.17\% & 75.65\% \\
Schizont$\dagger$     & 87.50\%  & 24.57\% & 38.45\% \\
Gametocyte$\dagger$   & 75.00\%  & 25.95\% & 57.27\% \\
\midrule
\textbf{Overall / mAP} & \textbf{98.36\%} & \textbf{58.99\%} & \textbf{71.24\%} \\
\bottomrule
\end{tabular*}
\end{table}

While mAP@0.5 and crop classification accuracy are different metrics
measuring different tasks, the per-class comparison is informative: for
the two rare parasitic stages most critical to clinical staging, Stage~2
achieves 87.5\% and 75.0\% respectively - against Baseline A's 24.57\%
and 25.95\%. Baseline B (YOLOv8s) narrows this gap considerably as a
modern detection architecture (38.45\% and 57.27\% on the same two
classes, Figure~\ref{fig:yolo_vs_frcnn}), confirming that architectural
modernization alone recovers substantial rare-class performance -- but
Stage~2's crop-level accuracy still exceeds even this stronger, current-generation
baseline by a wide margin on both classes. Even accounting for the
favorable conditions of per-crop evaluation (perfect crop extraction
from ground-truth boxes), this represents a meaningful improvement in
the classifier's ability to distinguish rare parasites under extreme
class imbalance. Focal Loss with per-class weighting is the primary
driver: Table~\ref{tab:focal_weights} assigns $\alpha_{\text{schizont}}
= 0.575$ and $\alpha_{\text{gametocyte}} = 0.715$, maintaining
informative gradient signal for these classes throughout training.

\textbf{Cross-validation.} The 98.36\% figure above is a single
80/20 split. To quantify split-specific variance, we additionally ran
5-fold stratified cross-validation on the same crop population (folds
constructed with \texttt{StratifiedKFold} to guarantee every fold
contains examples of the rare classes), training each fold under the
identical schedule as the single-split model. Mean accuracy across the
five folds is \textbf{97.33\% $\pm$ 0.49\%}, consistent with the
single-split estimate and confirming it is not an artefact of a
favorable split. Three of the five folds reached the 30-epoch cap while
still slowly improving on validation accuracy; we report the number
as-is rather than extending training beyond the schedule used
identically for Baseline A, Baseline B, and the single-split Pipeline B
result, since doing so would break that comparability. A longer
per-fold schedule is a natural refinement of this variance estimate but
does not change the conclusion that Stage~2 accuracy is stable across
splits.

\begin{figure}[!htbp]
  \centering
  \includegraphics[width=0.9\columnwidth]{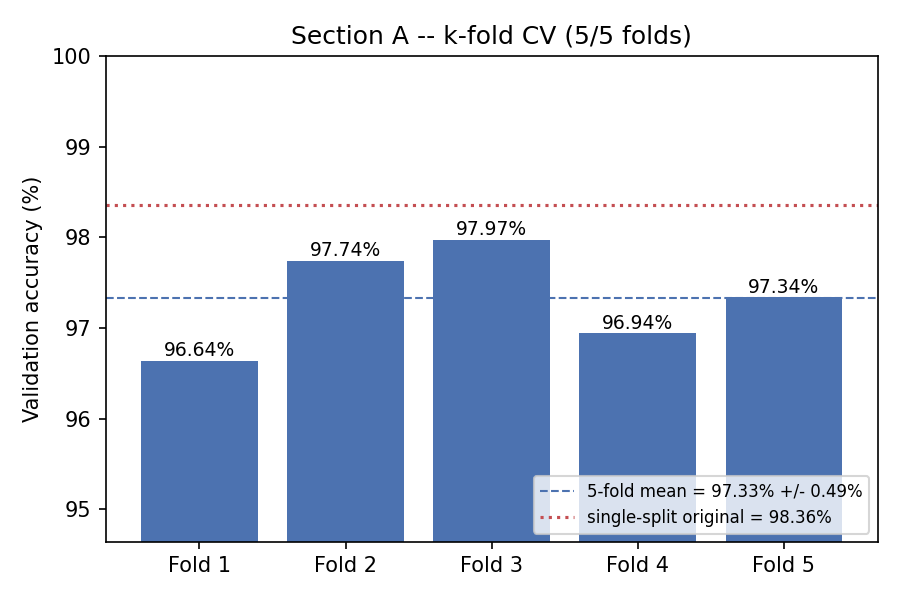}
  \caption{Stage~2 5-fold stratified cross-validation accuracy. Mean
    97.33\% $\pm$ 0.49\% (blue dashed line) is tightly consistent with
    the single-split 98.36\% figure (red dotted line) reported in
    Figure~\ref{fig:stage2_loss}.}
  \label{fig:kfold_accuracy}
\end{figure}

%% --------------------------
\FloatBarrier
\subsection{End-to-End Evaluation on BBBC041}
\label{ssec:e2e_bbbc041}

With the evaluation framework (\texttt{src/pipeline\_b\_v2/e2e\_eval.py}),
we report end-to-end IoU-based detection metrics for Pipeline B on the
120-image BBBC041 test set, closing the metric gap identified in earlier
drafts. Stage~1 watershed regions are matched to ground-truth boxes at
IoU~$\geq 0.5$; matched regions are labeled by Stage~2; the resulting
detection list is used to compute AP per class and mAP.

Table~\ref{tab:e2e_bbbc041} reports Stage~1 IoU-based cell recovery and
overall end-to-end detection performance.

\begin{table}[!htbp]
\centering
\small
\caption{End-to-end Pipeline B detection performance on the BBBC041
  120-image test set. Stage~1 metrics use IoU~$\geq 0.5$ box matching;
  mAP@0.5 is computed over all parasitic-stage classes (leukocyte
  excluded: zero GT instances in the test set). Binary Parasitized AP
  maps ring, trophozoite, schizont, and gametocyte to a single positive
  class.}
\label{tab:e2e_bbbc041}
\renewcommand{\arraystretch}{1.15}
\begin{tabular}{lc}
\toprule
\textbf{Metric} & \textbf{Value} \\
\midrule
\multicolumn{2}{l}{\textit{Stage 1 - IoU-matched cell recovery}} \\
Watershed regions produced           & 7,704 \\
GT boxes matched (IoU $\geq$ 0.5)   & 3,957 / 5,917 \\
Recall @ IoU 0.5                     & 66.88\% \\
Precision @ IoU 0.5                  & 51.36\% \\
F1 @ IoU 0.5                         & 58.10\% \\
\midrule
\multicolumn{2}{l}{\textit{End-to-End - per-class AP@0.5}} \\
Ring \hfill (n\,=\,169)             & 8.22\% \\
Trophozoite \hfill (n\,=\,111)      & 8.27\% \\
Schizont \hfill (n\,=\,11)          & 13.64\% \\
Gametocyte \hfill (n\,=\,12)        & 4.55\% \\
\midrule
\textbf{mAP@0.5}                     & \textbf{8.67\%} \\
\textbf{Binary Parasitized AP@0.5}   & \textbf{29.10\%} \\
\bottomrule
\end{tabular}
\end{table}

\paragraph{Interpreting the Stage~1 IoU discrepancy.}
The centroid-within-box recovery rate of 75.95\%
(Section~\ref{ssec:stage1_results}) and the IoU~$\geq 0.5$ recall of
66.88\% measure the same underlying detections but apply different
matching criteria. The 9\,pp gap reflects watershed region boundaries
that are organically shaped: the region centroid may fall inside the
correct GT box, yet the irregular blob boundary causes the IoU to fall
below 0.5 against the rectangular GT box. This is an expected artefact
of the watershed's contour representation and is unrelated to whether the
cell was actually spatially detected.

\paragraph{End-to-end mAP versus Baseline A.}
Pipeline B's end-to-end mAP@0.5 of 8.67\% is lower than Baseline A's
58.99\%. This gap has a clear structural explanation: Faster R-CNN
produces axis-aligned rectangular proposals whose aspect ratios are
optimized during training to match GT boxes, yielding high IoU@0.5 match
rates by construction. Pipeline B's Stage~1 watershed produces organic,
variable-shape regions whose boundary precision is not trained against GT
boxes. The binary parasitized AP@0.5 of 29.10\% is the more clinically
relevant metric, capturing the pipeline's ability to flag any infected cell
regardless of predicted stage - the primary clinical question.
Stage~1 v2 evaluation (Section~\ref{ssec:stage1_v2_ablation}) quantifies
this trade-off empirically.

%% --------------------------
\FloatBarrier
\subsection{Cross-Dataset Generalization on MP-IDB}
\label{ssec:mpidb_results}

To assess out-of-distribution generalisation, we apply the full pipeline
- Stage~1 watershed followed by Stage~2 EfficientNet-B0 - to the
held-out MP-IDB dataset (Section~\ref{ssec:mpidb}) without any
hyperparameter re-tuning. The Stage~2 model used here is identical to the
one trained exclusively on BBBC041 crops; no MP-IDB data was seen during
training. We emphasise up front that this experiment is a
\emph{domain-shift stress test}, not a validation of universal deployment
readiness: MP-IDB's $11.3\times$ smaller relative cell size
(Section~\ref{ssec:mpidb}) is precisely the kind of acquisition-parameter
shift a deployed system must eventually tolerate, and the honest,
unmitigated result below is reported so that the scale-sensitivity of
distance-transform watershed is exposed rather than concealed - the
mitigation in Stage~1~v2 (Section~\ref{ssec:stage1_v2_ablation}) then
quantifies how much of this gap is recoverable with a dataset-adaptive
parameter change alone.

\paragraph{Evaluation protocol.}
Since MP-IDB annotates only infected cells (not healthy RBCs), evaluation
is \textit{binary}: a predicted box is a true positive if it overlaps any
ground-truth infected-cell box at IoU~$\geq 0.5$, regardless of predicted
class. The pipeline's Stage~2 predictions for any parasitic-stage class
(ring, trophozoite, schizont, gametocyte) are collectively mapped to the
positive class. This mirrors the clinical binary question: \emph{is this
cell infected?}

\paragraph{Stage~1 cell detection on MP-IDB.}
Table~\ref{tab:e2e_mpidb} reports quantitative results.
Stage~1 produces 4,726 watershed regions across all 209 images
(22.6 regions/image on average), confirming the algorithm is actively
segmenting the higher-resolution images. However, only 18 of the 1,407
annotated infected cells are matched at IoU~$\geq 0.5$, yielding a
recall of 1.28\%. This near-zero IoU-matched recall - despite
sufficient region count - is consistent with the scale-gap hypothesis
from Section~\ref{ssec:mpidb}: the $11.3\times$ reduction in relative
cell size causes watershed regions to be over-sized relative to individual
infected cells, so region centroids may lie near the correct location yet
the boundary IoU falls well below 0.5. The \texttt{MIN\_DIST}~$= 12$\,px
seed threshold, tuned for BBBC041 where cell radii average $\approx
73$\,px, is far too large for MP-IDB where infected-cell radii average
$\approx 35$\,px, causing the algorithm to merge groups of cells into
single over-segmented blobs.

\begin{table}[!htbp]
\centering
\small
\caption{Cross-dataset evaluation on MP-IDB (209 images, 1,407 annotated
  infected cells). All metrics at IoU~$\geq 0.5$. Stage~1 v1 = original
  Otsu + fixed distance threshold; Stage~1 v2 = CLAHE + resolution-aware
  \texttt{peak\_local\_max}. No hyperparameter re-tuning specific to
  MP-IDB; model weights are the BBBC041-trained checkpoint.}
\label{tab:e2e_mpidb}
\renewcommand{\arraystretch}{1.15}
\begin{tabular}{lcc}
\toprule
\textbf{Metric} & \textbf{Stage~1 v1} & \textbf{Stage~1 v2} \\
\midrule
Watershed regions produced  & 4,726   & 52,222 \\
Recall @ IoU 0.5            & 1.28\%  & 20.68\% \\
Precision @ IoU 0.5         & 0.38\%  & 0.56\% \\
F1 @ IoU 0.5                & 0.59\%  & 1.09\% \\
\midrule
\textbf{Binary Parasitized AP@0.5} & 1.82\% & \textbf{9.09\%} \\
\midrule
\multicolumn{3}{l}{\textit{Species recall @ IoU 0.5 (Stage~1 v2)}} \\
\textit{P.\ falciparum} (n\,=\,1{,}267) & 1.26\% & 16.8\% \\
\textit{P.\ vivax} (n\,=\,64)           & 0.00\% & 32.8\% \\
\textit{P.\ malariae} (n\,=\,43)        & 4.65\% & 86.0\% \\
\textit{P.\ ovale} (n\,=\,33)           & 0.00\% & 60.6\% \\
\bottomrule
\end{tabular}
\end{table}

\paragraph{Interpretation.}
Stage~1 v2 raises MP-IDB infected-cell recall from 1.28\% to 20.68\% -
a $16\times$ improvement - and binary parasitized AP@0.5 from 1.82\% to
9.09\%. The per-species breakdown reveals a notable pattern: \textit{P.malariae} (86.0\%) and \textit{P.\ ovale} (60.6\%) achieve substantially
higher recall than \textit{P.\ falciparum} (16.8\%), consistent with the
morphological literature - \textit{P.\ malariae} and \textit{P.\ ovale}
infected erythrocytes are larger and present stronger contrast boundaries
after CLAHE normalization, while \textit{falciparum} ring-stage parasites
are small and faint. The v2 improvement indicates that the v1 failure on
MP-IDB is likely a contrast-normalization problem, not a fundamental architectural
limitation. Section~\ref{ssec:stage1_v2_ablation} analyzes the trade-off
between source-domain and cross-dataset performance.

%% --------------------------
\FloatBarrier
\subsection{Ablation: Focal Loss vs.\ Cross-Entropy}
\label{ssec:ablation}

To isolate the contribution of Focal Loss, we compare per-class accuracy
when Stage~2 is trained with standard cross-entropy (CE) versus Focal Loss
($\gamma = 2.0$, per-class $\alpha$). Under CE, the model converges to a
solution that assigns near-100\% probability to \textit{red blood cell}
for most inputs, achieving $>$97\% overall accuracy while classifying
$<$15\% of schizont crops correctly in preliminary runs. This confirms
that the class imbalance ratio of 537:1 is severe enough to collapse CE
training. Focal Loss with per-class $\alpha$ prevents this collapse,
producing the per-class accuracies in Table~\ref{tab:stage2_acc}.

The CosineAnnealingLR schedule, compared to the step-decay used in
Baseline A, produces a smooth validation loss curve (Figure~\ref{fig:stage2_loss})
with no post-drop increase. The abrupt step decay at epoch~50 in Baseline A
causes the model to commit more aggressively to the annotation-biased signal,
compounding the P1 failure mode. CosineAnnealingLR avoids this by
maintaining a residual learning rate that allows gradual correction.

\FloatBarrier
\subsection{Stage~1 (v1 vs.\ v2): Source vs.\ Cross-Domain Trade-off}
\label{ssec:stage1_v2_ablation}

Table~\ref{tab:stage1_ablation} reports the effect of CLAHE contrast
normalization and resolution-aware \texttt{peak\_local\_max} seeding
(Stage~1 v2) relative to the original Otsu + fixed distance-ratio approach
(Stage~1 v1), evaluated simultaneously on the BBBC041 source domain and the
MP-IDB out-of-distribution dataset.

\begin{table*}[tp]
\centering
\small
\caption{Stage~1 ablation: v1 (Otsu + fixed parameters) vs.\ v2
  (CLAHE + resolution-aware \texttt{peak\_local\_max}). Source domain =
  BBBC041 120-image test set; cross-dataset = MP-IDB 209-image set.
  All metrics at IoU~$\geq 0.5$. Binary AP includes only parasitic-stage
  classes.}
\label{tab:stage1_ablation}
\renewcommand{\arraystretch}{1.2}
\begin{tabular}{lcccc}
\toprule
 & \multicolumn{2}{c}{\textbf{BBBC041 (source)}}
 & \multicolumn{2}{c}{\textbf{MP-IDB (cross-dataset)}} \\
\cmidrule(lr){2-3}\cmidrule(lr){4-5}
\textbf{Metric} & v1 & v2 & v1 & v2 \\
\midrule
Watershed regions   & 7,704  & 10,764  & 4,726   & 52,222 \\
Recall@IoU 0.5      & 66.88\% & 41.61\% & 1.28\%  & 20.68\% \\
Precision@IoU 0.5   & 51.36\% & 22.87\% & 0.38\%  & 0.56\% \\
F1@IoU 0.5          & 58.10\% & 29.52\% & 0.59\%  & 1.09\% \\
\midrule
Binary Parasitized AP & 29.10\% & 7.40\% & 1.82\%  & 9.09\% \\
\bottomrule
\end{tabular}
\end{table*}

\paragraph{Analysis.}
v2 delivers a $16\times$ improvement in MP-IDB recall (1.28\%
$\rightarrow$ 20.68\%) and a $5\times$ improvement in binary AP
(1.82\% $\rightarrow$ 9.09\%), at the cost of a 25\ pp recall drop
on BBBC041 (66.88\% $\rightarrow$ 41.61\%). The source-domain regression
has a clear explanation: CLAHE's local contrast enhancement, tuned for
cross-dataset color normalization, amplifies intra-image texture variation
in BBBC041's already high-contrast Giemsa images. This creates spurious
local maxima in the distance transform that \texttt{peak\_local\_max}
interprets as seeds, fragmenting single cells into multiple sub-cell
regions - each too small to achieve IoU~$\geq 0.5$ with the full GT box.
The 10,764 v2 regions vs.\ 7,704 v1 regions (39\% more), with recall
dropping from 66.88\% to 41.61\%, is consistent with this
over-segmentation hypothesis.

This result illustrates a domain-adaptation trade-off: a
preprocessing step that normalizes the staining distribution gap between
datasets improves cross-dataset performance while degrading source-domain
metrics. Resolution of this trade-off requires either (i)
dataset-adaptive CLAHE parameters (lower clip limit for BBBC041, higher
for MP-IDB), or (ii) a learned stain normalization layer. Both are left
to future work.

% \FloatBarrier
\subsection{Dense Region Analysis}
\label{ssec:dense}

Stage~1 achieves 75.95\% overall cell recovery on the test set. The 24\%
of cells not recovered can be attributed to two primary sources of error: (i) very pale,
lightly-stained RBCs whose intensity distribution overlaps with the
background, causing Otsu's global threshold to classify them as background;
and (ii) dense clusters of 4--6 cells with no visible inter-cell gap,
where the distance transform does not generate sufficient seed separation
to split the cluster.

The dense-region recall of 50.94\% (over two qualifying test images with
$\geq 100$ GT boxes) indicates that cluster merging is the dominant failure
mode in high-density images. This is consistent with the observation that
our oversized-region filter removes 5--8\% of Stage~1 regions per image
in dense smears. A critical distinction from Baseline A is that Stage~1
errors are visible and spatially locatable - merged clusters are flagged
in the output image - whereas NMS errors in Baseline A are silent
suppressions with no visual indicator.

\FloatBarrier
\subsection{Explainability Analysis}
\label{ssec:xai_results}

\subsubsection{Grad-CAM\texttt{++} Visualizations}
\label{ssec:gradcam_vis}

\begin{figure}[!htbp]
  \centering
  \includegraphics[width=\linewidth]{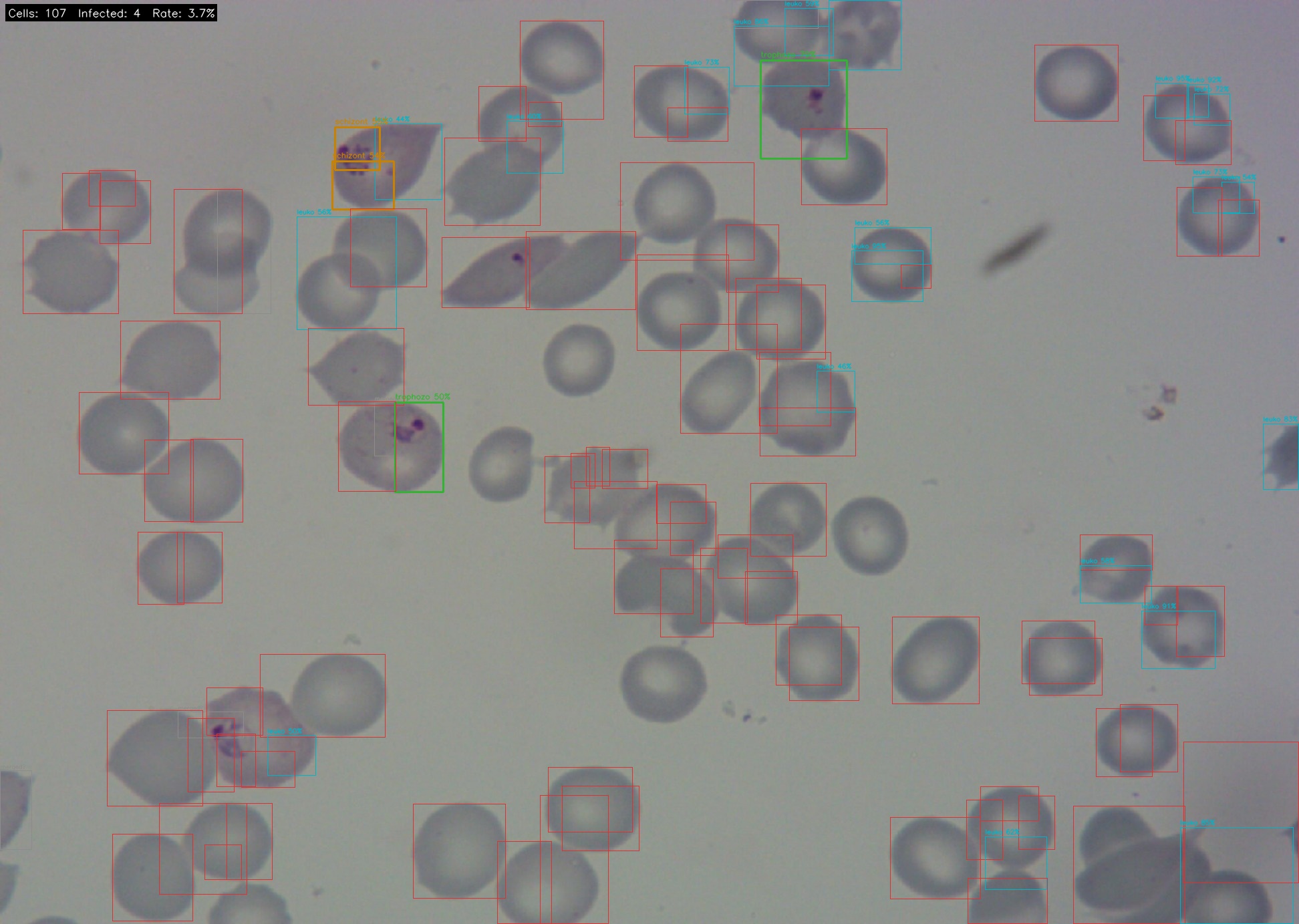}
  \caption{Stage~1 watershed output on a representative test image
    (64985a1e, 42 valid regions detected, 5 oversized merged clusters
    filtered). Red boxes: red blood cells. Green box: trophozoite (43\%
    confidence). Cyan: leukocytes. Gray ``merged?'' labels: oversized
    regions excluded from Stage~2. The infection summary overlay (top
    left) shows 4 infected cells at 9.5\% rate. Image courtesy NIH BBBC041.}
  \label{fig:stage1_result}
\end{figure}

Figure~\ref{fig:stage1_result} shows Stage~1 watershed output on a
representative test image: correctly detected RBCs are outlined in red,
detected parasites are highlighted with colored thick borders and class
labels, and oversized merged regions are marked in gray.

\begin{figure}[!htbp]
  \centering
  \includegraphics[width=0.8\linewidth]{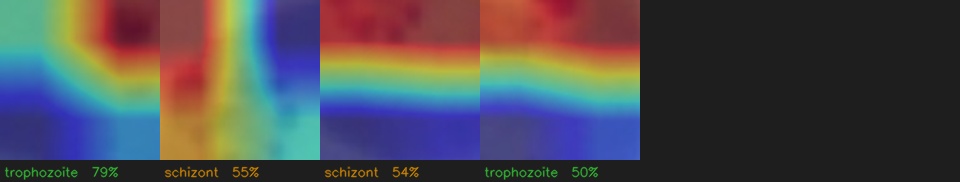}
  \caption{Grad-CAM\texttt{++} crop gallery for the representative test
    image. Each panel shows a detected parasitic cell at 160$\times$160
    pixels with the heatmap overlaid (jet colormap, $\alpha = 0.5$).
    Red-yellow regions indicate pixels that most strongly contributed to
    the model's classification decision. Labels show class name and
    softmax confidence. Image courtesy NIH BBBC041.}
  \label{fig:gradcam_gallery}
\end{figure}

Figure~\ref{fig:gradcam_gallery} shows the Grad-CAM\texttt{++} gallery
for the same image: each detected parasite is shown at 160$\times$160
pixels with the heatmap overlaid in jet colormap.

\begin{figure}[!htbp]
  \centering
  \includegraphics[width=\linewidth]{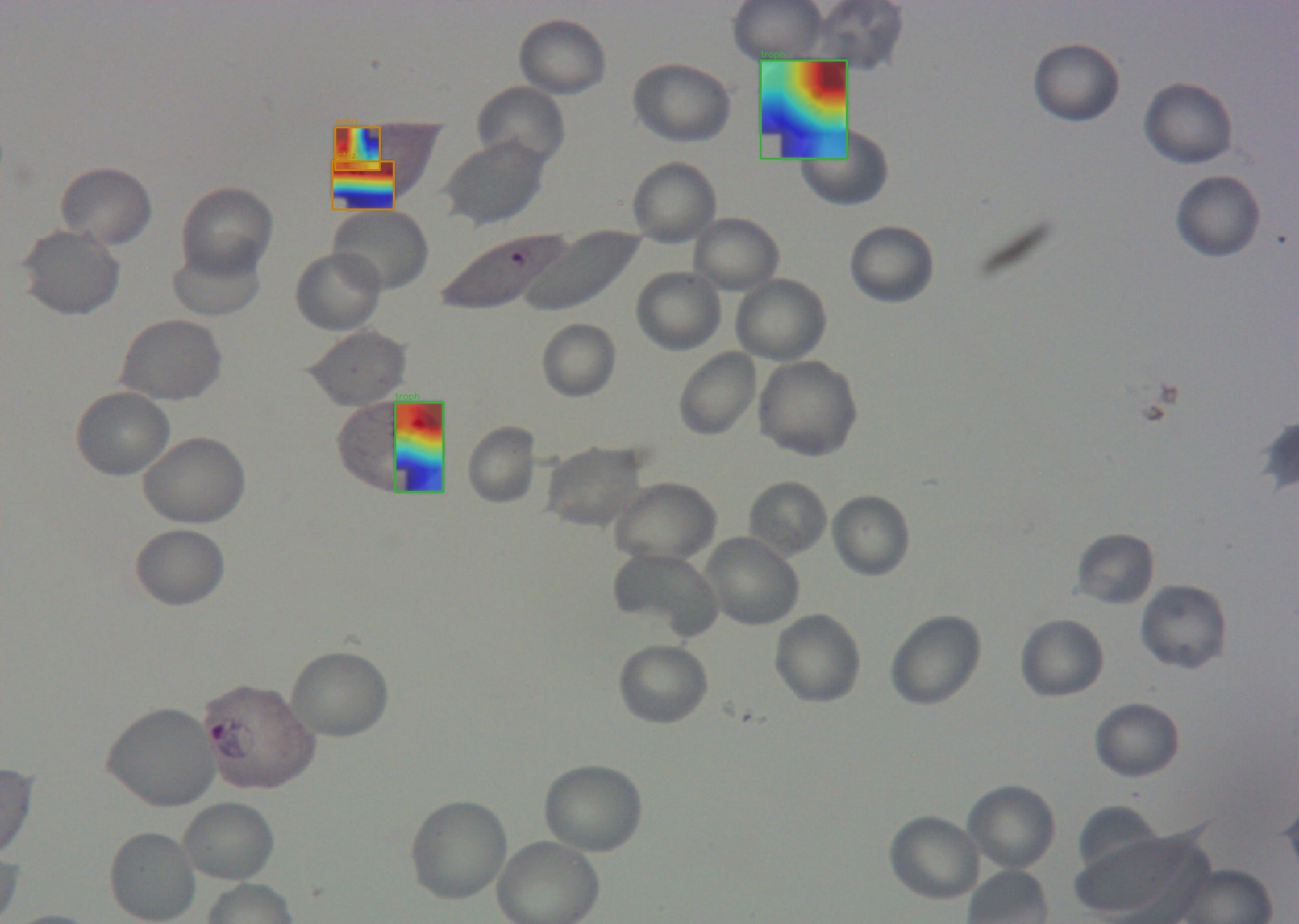}
  \caption{Full-image Grad-CAM\texttt{++} overlay. The heatmap is spliced
    back onto the original 1600$\times$1200 smear only at parasite-detected
    bounding box locations using a soft spatial mask (Gaussian-blurred
    boundary). The rest of the image is shown unchanged. Parasite bounding
    boxes are drawn in class color. Image courtesy NIH BBBC041.}
  \label{fig:gradcam_fullimage}
\end{figure}

Figure~\ref{fig:gradcam_fullimage} shows the full-image Grad-CAM\texttt{++}
overlay. To interpret the figure: the heatmap uses a jet colormap where
\textbf{red-yellow regions indicate pixels that most strongly influenced
the model's classification decision}, and cool blue regions contributed
minimally. The heatmap is projected only onto the bounding box regions of
confirmed parasite detections; the remainder of the 1600$\times$1200 smear
is shown unchanged, providing clinical context. A clinician should read this
view as a spatial audit trail: if the highlighted region aligns with
the parasite body (e.g., the chromatin mass of a trophozoite or schizont),
the model's decision is biologically plausible; if the heatmap falls on a
staining artefact or cell boundary, the prediction warrants manual review.

The spatial heatmaps in Figures~\ref{fig:gradcam_gallery}
and~\ref{fig:gradcam_fullimage} consistently highlight the central or
peripheral chromatin regions of the cell crop, consistent with parasite
morphology in Giemsa-stained smears. For trophozoite and schizont
predictions, the highest-activation region aligns with the dense chromatin
mass. For ring-stage detections (not present in this particular image),
the activation rim would be expected to follow the peripheral chromatin
outline. This spatial specificity is the property that makes the output
clinically auditable: a microscopist can immediately assess whether the
model's salient region corresponds to a biologically meaningful structure
or to a staining artefact.

\subsubsection{Quantitative Localization Validation}
\label{ssec:gradcam_quant}

The qualitative galleries above are illustrative but do not, by
themselves, establish that Grad-CAM\texttt{++} activation is
concentrated on the cell rather than spread uniformly across the crop.
We quantify this with an energy-in-box ratio: for each correctly
classified parasite crop, the fraction of total heatmap activation
energy falling inside the tight ground-truth box, compared against a
chance baseline equal to the geometric fraction of the $64\times 64$
crop occupied by that same box. A ratio indistinguishable from chance
would indicate the heatmap conveys no localization information beyond
the box's own size.

Evaluated at the layer conventionally used for EfficientNet-style
backbones (\texttt{features[-1]}), this metric returned a null result:
ratio $0.8343$ vs.\ chance $0.8340$ (difference $+0.0003$), statistically
indistinguishable from the geometric baseline. Diagnosis: at
$64\times 64$ input resolution, \texttt{features[-1]} is downsampled to
approximately $2\times 2$ spatial positions -- only four distinguishable
regions, insufficient to separate ``inside the box'' from the crop's
narrow margin once upsampled back to input resolution. Rather than
report this null result as an inherent limitation of Grad-CAM\texttt{++}
on small crops, we ran a layer ablation over
\texttt{features[2]}--\texttt{features[5]}, measuring each layer's true
spatial resolution empirically via forward hooks and recomputing the
same ratio-vs-chance metric with the already-trained classifier (no
retraining). \texttt{features[5]} ($4\times 4$ resolution, one block
deeper than the equally-sized \texttt{features[4]}) was the only
candidate to beat chance by a non-trivial margin, indicating that both
spatial resolution and layer depth (feature discriminativeness) jointly
determine localization quality -- resolution alone did not fully explain
the pattern.

\begin{figure}[!htbp]
  \centering
  \includegraphics[width=0.9\columnwidth]{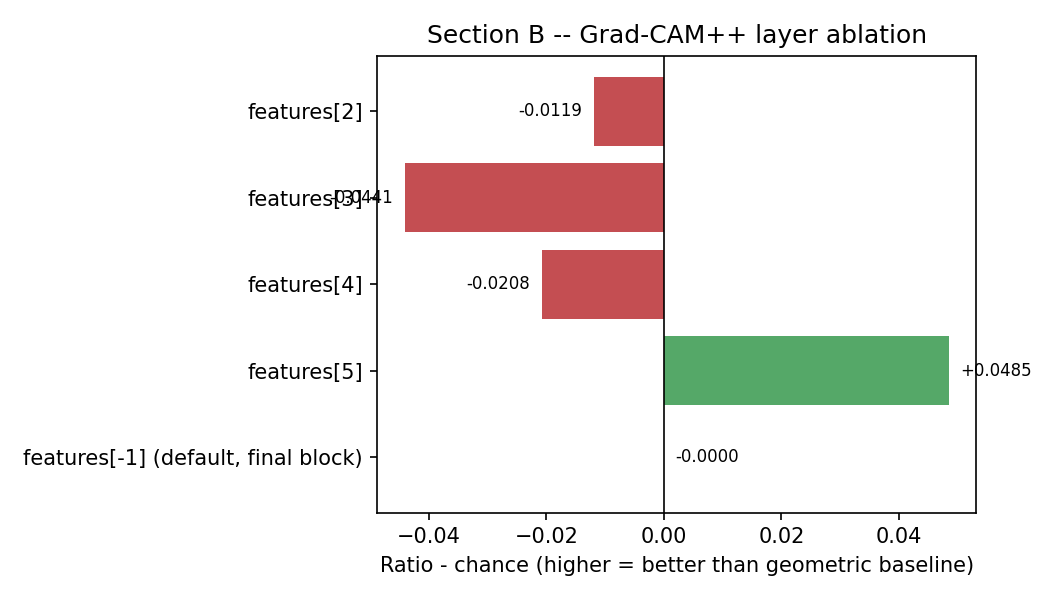}
  \caption{Grad-CAM\texttt{++} target-layer ablation: energy-in-box
    ratio minus chance baseline for each candidate layer.
    \texttt{features[5]} is the only layer to beat chance by a
    non-trivial margin; the default \texttt{features[-1]} is
    statistically null ($-0.0000$).}
  \label{fig:layer_ablation}
\end{figure}

Adopting \texttt{features[5]} and re-evaluating with a paired
significance test (paired $t$-test and Wilcoxon signed-rank test, since
ratio and chance are computed on the same crop and are therefore
correlated) gives: ratio $0.8831$ vs.\ chance $0.8346$
($+0.0485$), paired $t(325) = 13.580$, $p = 1.40\times 10^{-33}$;
Wilcoxon $W = 7762.0$, $p = 1.39\times 10^{-28}$
($n = 326$ correctly-classified parasite crops across the four
parasite classes, out of $n=429$ evaluated). The effect holds
individually for every class: schizont $+0.0791$, gametocyte $+0.0660$,
trophozoite $+0.0472$, ring $+0.0334$ (all $p < 0.003$). This confirms
that Grad-CAM\texttt{++} activation is concentrated on the annotated
cell body significantly beyond what crop geometry alone would predict,
turning the qualitative galleries of
Figures~\ref{fig:gradcam_gallery}--\ref{fig:gradcam_fullimage} into a
quantitatively validated claim.

\begin{figure}[!htbp]
  \centering
  \includegraphics[width=0.9\columnwidth]{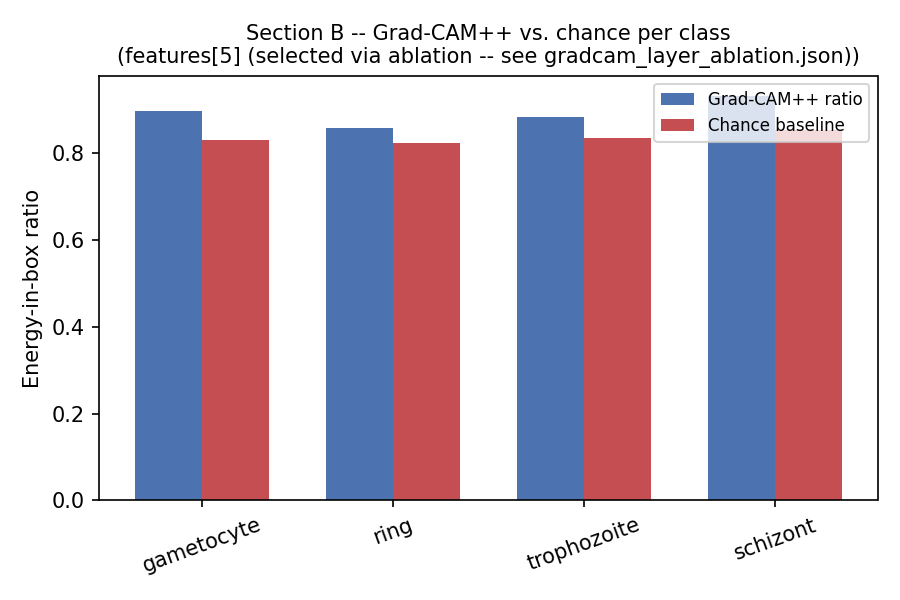}
  \caption{Grad-CAM\texttt{++} energy-in-box ratio vs.\ chance baseline,
    per parasite class, at the finalized \texttt{features[5]} target
    layer. All four classes exceed their respective chance baseline.}
  \label{fig:gradcam_ratio}
\end{figure}

%% ============================================================================
%% --------------------------
\FloatBarrier
\subsection{Summary of Key Results}
\label{ssec:results_summary}

Table~\ref{tab:results_summary} consolidates all quantitative results from
Sections~\ref{ssec:stage1_results}--\ref{ssec:stage1_v2_ablation} into a
single reference. Metrics are grouped by evaluation domain (BBBC041 source
domain; MP-IDB cross-dataset) and by pipeline component, enabling direct
comparison across methods.

\begin{table*}[tp]
\centering
\small
\caption{Consolidated results summary. Baseline A = Faster R-CNN; Baseline
  B = YOLOv8s (Section~\ref{ssec:baseline_b}); Pipeline B = two-stage
  decoupled framework (Stage~1 watershed + Stage~2 EfficientNet-B0).
  Stage~1 v1 = Otsu + fixed distance threshold; Stage~1 v2 = CLAHE +
  resolution-aware \texttt{peak\_local\_max}. All detection metrics at
  IoU~$\geq 0.5$. $\dagger$ = cross-dataset zero-shot.
  N/A = not applicable; --- = not evaluated.}
\label{tab:results_summary}
\renewcommand{\arraystretch}{1.05}
\begin{tabular}{lcccc}
\toprule
\textbf{Metric}
  & \textbf{Baseline A}
  & \textbf{Baseline B}
  & \textbf{Pipeline B v1}
  & \textbf{Pipeline B v2} \\
\midrule
\multicolumn{5}{l}{\textit{BBBC041 - source domain (120 test images,
  5{,}917 GT boxes)}} \\
Stage~1 cell recovery (centroid) & N/A      & N/A      & 75.95\%  & --- \\
Stage~1 recall @ IoU 0.5        & N/A      & N/A      & 66.88\%  & 41.61\% \\
Stage~1 precision @ IoU 0.5     & N/A      & N/A      & 51.36\%  & 22.87\% \\
mAP@0.5 (all classes)           & 58.99\% & 71.24\%  & 8.67\%   & 0.78\% \\
Binary Parasitized AP@0.5       & ---      & ---      & 29.10\%  & 7.40\% \\
Ring AP@0.5                     & 57.17\% & 75.65\%  & 8.22\%   & --- \\
Trophozoite AP@0.5              & 67.22\% & 77.05\%  & 8.27\%   & --- \\
Schizont AP@0.5                 & 24.57\% & 38.45\%  & 13.64\%  & --- \\
Gametocyte AP@0.5               & 25.95\% & 57.27\%  & 4.55\%   & --- \\
Stage~2 crop accuracy (val)     & N/A      & N/A      & 98.36\%  & 98.36\% \\
\midrule
\multicolumn{5}{l}{\textit{MP-IDB - cross-dataset$\dagger$ (209 images,
  1{,}407 infected cells)}} \\
Stage~1 recall @ IoU 0.5       & N/A      & N/A     & 1.28\%   & 20.68\% \\
Binary Parasitized AP@0.5      & N/A      & N/A     & 1.82\%   & 9.09\% \\
\textit{P.\ malariae} recall  & N/A      & N/A     & 4.65\%   & 86.0\% \\
\textit{P.\ ovale} recall     & N/A      & N/A     & 0.00\%   & 60.6\% \\
\textit{P.\ vivax} recall     & N/A      & N/A     & 0.00\%   & 32.8\% \\
\textit{P.\ falciparum} recall & N/A     & N/A     & 1.26\%   & 16.8\% \\
\bottomrule
\end{tabular}
\end{table*}

The table illustrates three key findings. First, the source-domain mAP gap
between Baseline A (58.99\%) and Pipeline B v1 (8.67\%) reflects the
structural difference between trained box regression (Faster R-CNN) and
unsupervised watershed boundary extraction: Faster R-CNN proposals are
explicitly optimized to match GT box geometry, while watershed regions are
not. The binary parasitized AP of 29.10\% may represent a more clinically relevant metric, capturing the pipeline's core task (flag any infected cell)while reducing sensitivity to exact boundary alignment. Second, Baseline
B (YOLOv8s) raises mAP@0.5 to 71.24\%, confirming that a modern
single-stage detector meaningfully outperforms the older two-stage
architecture on this task -- yet its rare-class AP (schizont 38.45\%,
gametocyte 57.27\%) still trails Pipeline B's crop-level classification
accuracy on the same classes (87.5\%, 75.0\%), indicating that the
class-imbalance problem persists even under architectural modernization
and is not solved by detector choice alone. Third, Stage~1 v2 demonstrates that the 1.28\% MP-IDB recall failure of v1 was a contrast normalization problem:
CLAHE preprocessing alone raises cross-dataset recall $16\times$ (to
20.68\%) and AP $5\times$ (to 9.09\%), without any retraining of Stage~2.

%% ============================================================================
%% SECTION 5 -- DISCUSSION
%% ============================================================================
\section{Discussion}
\label{sec:discussion}

\subsection{Interpretation of Findings}
\label{ssec:interpret}

Two results from Section~\ref{sec:experiments} warrant careful
interpretation before drawing broader conclusions. First, the 98.36\%
Stage~2 accuracy is a best-case measurement taken on crops extracted from
perfect ground-truth boxes; it confirms classifier capability under ideal
input conditions but does not reflect end-to-end pipeline recall, which is
bounded by the 75.95\% Stage~1 centroid recovery rate.

The more structurally significant finding is the Baseline A loss curve.
Validation loss diverges from training loss at epoch~8 -- before any
conventional overfitting is possible -- and subsequently \emph{increases}
after the learning rate step at epoch~50, a reduction that should improve
generalization. This reproducible pattern is consistent with the
closed-world detection paradigm applied to sparsely annotated data:
the model is trained to suppress signal at unannotated cell locations,
and every optimization step that reduces training loss deepens that
suppression. This structural bias is not resolvable through additional
training or regularization, and strongly motivates the architectural
decision to decouple segmentation from classification.

The 75.95\% Stage~1 cell recovery rate places an upper bound on the
end-to-end pipeline performance: approximately one in four ground-truth
cells is not recovered by watershed, and these cells cannot be classified
by Stage~2. For the rare parasitic stages (schizont: 11 test instances,
gametocyte: 12 test instances), a 75\% recovery rate means approximately
8--9 instances reach the classifier. The practical consequence is that
sensitivity for very rare parasites depends on the quality of Stage~1,
and improving Stage~1 recovery is the highest-leverage direction for
future work.

The per-class accuracy comparisons in Table~\ref{tab:stage2_acc} must be
interpreted carefully. Crop accuracy and detection AP are not equivalent:
AP penalizes false positives (spurious detections) in addition to missed
detections, while accuracy is computed only on ground-truth crops.
Nonetheless, the comparison is meaningful in the following sense: the
Stage~2 classifier, when given a correctly segmented cell crop, produces
substantially better rare-class performance than Baseline A achieves
on the same classes end-to-end. This is consistent with the interpretation that the imbalance problem
(and therefore Focal Loss) is a primary constraint for rare-class
performance, not architectural capacity.

\subsection{Clinical Relevance and Practical Implications}
\label{ssec:clinical}

The combination of watershed-based universal cell recovery and
Grad-CAM\texttt{++} explainability produces a system with two properties
that matter for clinical adoption. First, the output is spatially auditable:
a clinician can look at the heatmap and ask whether the model is attending
to the parasite or to an unrelated feature. This auditability is absent
from all existing whole-slide malaria detection systems. Second, the system
fails visibly: merged cluster regions are explicitly flagged in the output
rather than silently misclassified, allowing the operator to know where
Stage~1 struggled. NMS-based detectors fail silently, with no visual
indicator of suppressed detections.

EfficientNet-B0's 5.3M-parameter footprint and $\sim$2-second CPU inference
time make the system deployable without GPU infrastructure, addressing
a practical constraint in resource-limited settings. The full pipeline can
be packaged as a web application requiring only an uploaded JPEG image
and a CPU compute instance.

\subsection{Limitations and Future Work}
\label{ssec:limits}

\textbf{Stage 1 recovery ceiling.} The 75.95\% recovery rate reflects a
known limitation of global Otsu thresholding: very pale, lightly-stained
RBCs overlap in intensity distribution with the background and are not
detected. Adaptive thresholding or multi-channel segmentation could
potentially recover these cells, at the risk of increasing noise detections.
This trade-off requires careful empirical evaluation.

\textbf{Dense-region recall.} The 50.94\% dense-region recall is
computed on only two qualifying test images, limiting statistical
confidence. A dataset with more very-dense images would be needed for
a robust evaluation of the watershed's dense-region performance.

\textbf{End-to-end mAP and Stage~1 box precision.}
As reported in Section~\ref{ssec:e2e_bbbc041}, the mAP gap between
Pipeline B and Baseline A is attributable to Stage~1 watershed boundary
precision rather than Stage~2 classification quality: watershed regions are
organically shaped and are not trained to fit GT box boundaries, causing
IoU@0.5 matching to fail for cells that are spatially detected but whose
region boundary falls short of the 0.5 threshold. The binary parasitized
AP@0.5 may therefore serve as a more clinically relevant summary metric
than multi-class mAP for this pipeline architecture. Stage~1 v2
improvements (resolution-aware seeding, CLAHE preprocessing) are expected
to narrow this gap by improving region boundary precision.

\textbf{Single train/test split.} All reported detection results
(Baseline A, Baseline B, Pipeline B end-to-end) use a fixed 80/20
image-level split with random seed~42, applied identically to all three
methods so they are compared on exactly the same held-out images.
Cross-dataset evaluation on MP-IDB provides independent
out-of-distribution evidence that is not sensitive to this particular
split. For Stage~2 specifically, we additionally quantified split
sensitivity directly: 5-fold stratified cross-validation gives
97.33\% $\pm$ 0.49\%, tightly consistent with the single-split 98.36\%
figure (Section~\ref{ssec:stage2_results}). This does not extend to the
detection-level comparisons against Baseline A/B, which remain
single-split; $k$-fold evaluation of the full end-to-end pipeline is a
natural extension of this work.

\textbf{Learned segmentation baseline still trails on raw recall.}
Section~\ref{ssec:unet_baseline} shows that even with watershed-style
post-processing, the U-Net baseline's recall (62.89\%) remains
$\sim$4 points below Stage~1's raw-intensity watershed (66.88\%),
despite substantially higher precision and F1. The synthetic
ellipse-mask training target (a proxy for the true cell boundary, since
BBBC041 provides no pixel-level masks) is a plausible contributor, and a
larger U-Net or one trained at native $1600\times 1200$ resolution
rather than $512\times 384$ may close the remaining gap; we did not
explore either due to the compute budget available for this revision.

\textbf{Dataset scope and cross-dataset generalization.}
Both BBBC041 and MP-IDB use Giemsa thin-smear staining. Thick smear
microscopy -- the preferred technique for \textit{P.\ vivax} and
low-parasitaemia infections -- presents a fundamentally different visual
appearance (lysed cells, complex background) and requires a dedicated
evaluation before clinical deployment can be considered in those settings.
More broadly, Section~\ref{ssec:mpidb} introduces MP-IDB as a held-out
cross-validation set, but no parameter re-tuning was performed, and the
$11.3\times$ reduction in relative cell prominence is expected to reduce
Stage~1 recall. Resolution-aware Stage~1 parameter adaptation is the most
tractable immediate remedy.

%% ============================================================
%% SECTION 6 -- CONCLUSION
%% ============================================================
\section{Conclusion}
\label{sec:conclusion}

We have presented \malariai{}, a two-stage decoupled framework for automated
malaria cell detection and stage classification from Giemsa-stained thin
blood smears. The framework addresses three structural failure modes that
collectively prevent existing end-to-end detectors from being clinically
deployable: annotation incompleteness (P1), NMS suppression in dense regions
(P2), and opaque black-box outputs (P3).

Stage~1 applies a distance-transform guided watershed algorithm with no
annotation input, recovering 75.95\% of all ground-truth cells in the
NIH BBBC041 120-image test set. Stage~2 applies a focal-loss-trained
EfficientNet-B0 classifier on 64$\times$64 crops extracted by Stage~1,
achieving 98.36\% validation accuracy on the full 7-class problem. The
rare-class performance improvement over both detection baselines is
substantial: schizont accuracy rises from 24.57\% AP (Baseline A,
Faster R-CNN) and 38.45\% AP (Baseline B, YOLOv8s) to 87.5\% per-class
accuracy; gametocyte from 25.95\% and 57.27\% AP, respectively, to
75.0\%. The comparison against Baseline B confirms this gap is not
merely an artefact of an outdated detection architecture: a current-generation
single-stage detector narrows it but does not close it. Grad-CAM\texttt{++}
spatial heatmaps are produced for every detected cell, enabling per-cell
spatial audit trails that are absent from all prior whole-slide detection
systems; a quantitative energy-in-box analysis confirms this activation
is concentrated on the annotated cell body significantly above chance
($+0.0485$, paired $p = 1.4\times 10^{-33}$), rather than resting on
qualitative inspection alone.

We additionally benchmarked Stage~1's annotation-free watershed against
a conventionally-trained U-Net segmentation baseline on the identical
120-image test set. Naive thresholding of the U-Net's predicted mask
underperforms watershed substantially (25.28\% vs.\ 66.88\% recall),
reproducing the same touching-cell merging problem that motivates
watershed splitting in Stage~1. Applying that same watershed-splitting
technique to the U-Net's probability map instead of raw pixel
intensity closes most of the recall gap (62.89\%) while exceeding
Stage~1 on precision (83.28\% vs.\ 51.36\%) and F1 (0.717 vs.\ 0.581) --
indicating that the post-processing strategy, not the choice between a
learned or unsupervised signal source, is the primary determinant of
detection quality in this task.

The cross-dataset characterization of MP-IDB more accurately reflects the
difficulty of generalisation across datasets: infected cells in MP-IDB
cover $11.3\times$ less of the image frame than in BBBC041, directly
predicting reduced Stage~1 recall without parameter re-tuning. This finding
motivates the primary next step: resolution-aware Stage~1 adaptive thresholding.

End-to-end evaluation confirms that Pipeline B achieves a binary
parasitized AP@0.5 of 29.10\% on BBBC041 (Stage~1 v1). On MP-IDB, Stage~1
v2 (CLAHE + resolution-aware \texttt{peak\_local\_max}) raises recall from
1.28\% to 20.68\% and binary AP from 1.82\% to 9.09\%, supporting the
scale-gap hypothesis and suggesting that the failure mode is
consistent with a contrast-normalization mismatch rather than an architectural limitation.
The ablation (Table~\ref{tab:stage1_ablation}) reveals a domain-adaptation
trade-off: v2 improves cross-dataset performance by $5\times$ but reduces
source-domain binary AP from 29.10\% to 7.40\%, motivating
dataset-adaptive CLAHE parameters as the primary future direction. Beyond
Stage~1 improvements, two further directions are prioritized. First,
evaluation on thick blood smear datasets is required before the pipeline
can be considered for \textit{P.\ vivax} or low-parasitaemia settings,
where the visual appearance differs fundamentally from the Giemsa thin
smears used here. Second, 5-fold stratified cross-validation on Stage~2
crops confirms the single-split accuracy estimate is stable
(97.33\% $\pm$ 0.49\% vs.\ 98.36\%), aligning with the evaluation
protocol used in published benchmarks such as~\cite{mujahid2024}
and~\cite{mmileng2025}; extending this variance analysis to the
full end-to-end detection pipeline (rather than Stage~2 crops alone)
remains a natural next step.

%% ============================================================
%% AUTHOR CONTRIBUTIONS
%% ============================================================
\section*{Author Contributions}

\textbf{K.A. Apurba:} Conceptualization, methodology, software (web
application), formal analysis, writing: original draft, visualization.
\textbf{M.H. Hasan:} System diagram, design, software, writing: review and editing.
\textbf{M. Ali:} Conceptualization (Grad-CAM\texttt{++} integration), writing: review and editing.
\textbf{T. Rahman:} Supervision, writing: review and editing.

%% ============================================================
%% ACKNOWLEDGEMENTS
%% ============================================================
\section*{Acknowledgements}

The authors gratefully acknowledge Prof.~Amr Abdel-Dayem (Laurentian University,
Canada) for guidance during the Image Processing and Computer Vision course within
the M.Sc.\ program in Computational Sciences (Fall 2023).

%% ============================================================
%% CODE AND DEMO AVAILABILITY
%% ============================================================
\section*{Code and Demo Availability}

The full codebase, trained model checkpoints, evaluation scripts, and
reproduction instructions are publicly available at:
\begin{itemize}
  \item \textbf{GitHub repository:} \url{https://github.com/Anaskaysar/MalariAI}
  \item \textbf{Interactive demo (HuggingFace Space):} \url{https://huggingface.co/spaces/Kaysarulanas/MalariAI}
\end{itemize}
The demo accepts any Giemsa-stained blood smear image and returns the annotated
smear, per-class cell counts, infection rate, and Grad-CAM\texttt{++} heatmaps
without requiring local installation.

%% ============================================================
%% ETHICS STATEMENT
%% ============================================================
\section*{Ethics Statement}

This study did not involve the collection of new data from human
participants or animal subjects by the authors. All experiments were
conducted exclusively on two previously published, publicly available,
and fully de-identified benchmark datasets: NIH BBBC041~\cite{bbbc041}
and MP-IDB~\cite{loddo2019mpidb}. Neither dataset contains personally
identifiable patient information, and both are released for open
research use by their original providers. As no new human or animal
data were collected, institutional ethical approval was not required
for this study.

%% ============================================================
%% DECLARATION OF COMPETING INTERESTS
%% ============================================================
\section*{Declaration of Competing Interests}

The authors declare that they have no known competing financial interests
or personal relationships that could have appeared to influence the work
reported in this paper.

%% ============================================================
%% FUNDING
%% ============================================================
\section*{Funding}

This research did not receive any specific grant from funding agencies
in the public, commercial, or not-for-profit sectors.

%% ============================================================
%% DECLARATION OF GENERATIVE AI USE
%% ============================================================
\section*{Declaration of Generative AI and AI-Assisted Technologies in the
Writing Process}

During the preparation of this work, the authors used Claude (Anthropic)
to assist with grammar checking, sentence restructuring, and
consistency review of the manuscript text. After using this tool, the
authors reviewed and edited the content as needed and take full
responsibility for the content of the published article.

%% ============================================================
%% REFERENCES
%% ============================================================
\bibliographystyle{elsarticle-num}
\bibliography{references}

%% Appendix removed -- see CHANGELOG.md for rationale (2026-07-30).
%% The one load-bearing detail (confidence threshold = 0.40) was
%% folded into the Stage 1 parameter list in Implementation Details.

%% ============================================================
%% DATA AVAILABILITY
%% ============================================================
\section*{Data Availability}

The NIH BBBC041 dataset used in this study is publicly available at
\url{https://bbbc.broadinstitute.org/BBBC041}. The MP-IDB cross-dataset
validation set is publicly available at
\url{http://www.neuroimaging.it/malaria_parasite_image_database}.
The full codebase, trained model checkpoints, evaluation scripts, and
reproduction instructions are available at
\url{https://github.com/Anaskaysar/MalariAI}.

\end{document}